\documentclass[10pt]{article}

\usepackage[english]{babel}
\usepackage{amsmath}
\usepackage{amssymb}
\usepackage{bm}
\usepackage{bbm}
\usepackage{mathrsfs,color}
\usepackage{graphics,graphicx,theorem}
\usepackage{dsfont}
\usepackage{setspace}

\usepackage{mathtools}

\usepackage{natbib}
\usepackage{multirow}
\usepackage{hhline}
\usepackage{hyperref}

\usepackage{authblk}

\hypersetup{
	colorlinks=true,
	urlcolor=blue,
	citecolor=blue,
	linktoc=all,
	linkcolor=red} 

\usepackage[scriptsize]{subfigure}
\allowdisplaybreaks

\makeatletter
\renewcommand\@biblabel[1]{}
\makeatother

\def\bSig\mathbf{\Sigma}

\newcommand{\qed}{$\square$}

\newtheorem{lemma}{Lemma}[section]
\newtheorem{definition}{Definition}[section]
\newtheorem{proposition}{Proposition}[section]

\usepackage[top=1.2in,bottom=1.2in,left=1.2in,right=1.2in]{geometry}

\begin{document}

\title{\bf {\Large{An Information Theoretic approach to Post Randomization Methods under Differential Privacy}}}

\author[,1]{Fadhel Ayed \thanks{fadhel.ayed@gmail.com}}
\author[,2]{Marco Battiston \thanks{marco.battiston@stats.ox.ac.uk}}
\author[,3]{Federico Camerlenghi \thanks{federico.camerlenghi@unimib.it}}

\affil[1]{University of Oxford}
\affil[2]{Lancaster University}
\affil[3]{University of Milano--Bicocca}

\date{}
\maketitle
\thispagestyle{empty}

\setcounter{page}{1}

\maketitle

\begin{abstract}
Post Randomization Methods (PRAM) are among the most popular disclosure limitation techniques for both categorical and continuous data. In the categorical case, given a stochastic matrix $M$ and a specified variable, an individual belonging to category $i$ is changed to category $j$ with probability $M_{i,j}$. Every approach to choose the randomization matrix $M$  has to balance between two desiderata: 1) preserving as much statistical information from the raw data as possible; 2) guaranteeing the privacy of individuals in the dataset. This trade-off has generally been shown to be very challenging to solve. In this work, we use recent tools from the computer science literature and propose to choose $M$ as the solution of a constrained maximization problems. Specifically,  $M$ is chosen as the solution of a constrained maximization problem, where we maximize the Mutual Information between raw and transformed data, given the constraint that the transformation satisfies the notion of Differential Privacy. For the general Categorical model, it is shown how this maximization problem reduces to a convex linear programming and can be therefore solved with known optimization algorithms.
\end{abstract}


\section{Introduction}
\label{sec:1}

Data from census or survey studies are among the most useful sources of information for social and political studies. However, when statistical and governmental agencies release microdata to the public, they often encounter ethical and moral issues concerning the possible privacy leak for individuals present in the dataset. 
Anonymization techniques, like encrypting or removing personally identifiable information, have been widely used with the hope of ensuring  
privacy protection. However, recent studies by \cite{Gym13}, \cite{Hom08}, \cite{Nar08}, \cite{Swe97} 
have shown that, even after removing directly identifying variables, like names or national insurance numbers, the potential for breaches of confidentiality is still present. Specifically,
an intruder might still be able to identify individuals by cross-classifying categorical variables in the dataset and matching them with some external database. This kind of privacy problems have been widely considered in the statistical literature and different measures of \emph{disclosure risk} have been proposed to assess the riskiness of specific dataset.

Different disclosure limitation techniques have been proposed, like rounding, suppression of extreme values or entire variables, sampling or perturbation techniques. \emph{Post Randomization Methods} (PRAM) are among the most used techniques for disclosure risk limitation. See \cite{DeW97}, \cite{Gou97}, \cite{Koo97}. With these techniques, before releasing the dataset, the data curator randomly changes the values of some categorical identifying variables, like gender, job or age, of some individuals in the dataset. 
In a recent paper, \cite{Shl10} consider PRAM and random data swapping of a geographical variable and propose a way of computing measures of disclosure risk to assess whether these techniques have been effective in ``privatizing'' the dataset. The choice of the geographical variable is motivated by the fact that, by swapping or changing it, it is usually less likely to generate unreasonable combinations of categorical variables, like for instance a pregnant man or a 10 year old lawyer. In order to implement PRAM, they consider a stochastic matrix $M$, where the $(i,j)$ entry of this matrix gives the probability that an individual from location $i$ has his geographical variable swapped to location $j$. Given this known matrix $M$, \cite{Shl10} suggest some measures of risk and related estimation methods. However, an open problem is to decide how the data curator should actually choose the matrix $M$ in order to guarantee an effective level of privacy. 

Over the last ten years, a new approach to data protection, called \emph{Differential Privacy} (see \cite{Dwo06}), has become more and more popular in the computer science literature and has been implemented in their security protocols by IT companies 
(\cite{Ela15}, \cite{Erl14},\cite{Mac08}). This new framework finds its roots in the cryptography literature and prescribes to transform the original data, containing sensitive information, using a channel or \emph{mechanism} $Q$ into a sanitized dataset. The mechanism $Q$ should be chosen carefully, in such a way that, by only looking at the released dataset, an intruder will have very low probability of guessing correctly  
 the presence or absence of a specific individual in the raw data and, therefore, the privacy of the latter will be preserved. Differential Privacy formalizes mathematically this intuitive idea. We will provide a short review on it in Subsection \ref{sub:DiffPriv}. 

In this work, we bring together ideas from the Disclosure Risk and Differential Privacy literature to propose a formal way of choosing the stochastic matrix $M$ used in PRAM. Specifically, when choosing $M$, we need to balance two conflicting goals: 1) on the one hand, we want the application of $M$ to make the dataset somehow private; 2) on the other hand, we also want that the released dataset preserves as much statistical information as possible from the raw data. In order to balance this trade-off, we propose to choose $M$ as the solution of a constrained maximization problem. We maximize the Mutual Information between the released and the raw dataset, hence guaranteeing preservation of statistical information and achieving goal 2).  Mutual Information is a common measure of dependence between random variables used in probability and information theory.  In order to guarantee also goal 1), we introduce a constraint in the maximization problem by imposing that the application $M$ satisfies differential privacy, therefore the resulting mechanism based on $M$ can  formally be considered private. 
We show how this optimization problem results in a convex maximization problem under linear contraints and can therefore being solved efficiently by known optimization algorithms. 

The rest of this work is organized as follows. In Section 2, first we will briefly review the disclosure risk problem in Subsection \ref{sub:DiscRisk} and then the tools needed for our approach. Specifically, we review  Mutual Information in Subsection \ref{sub:MutInf} and   Differential Privacy in Subsection \ref{sub:DiffPriv}. In Section \ref{sec:3}, we formalize the proposed constrained maximization problem to choose the stochastic matrix $M$ in PRAM and show that this choice is made by solving a convex optimization problem under linear constraints. Section \ref{sec:4} contains a simulation study showing first the effect of the Diffential Privacy constraint on simulated data and  then the effect of different choices of $M$ using a real dataset of a survey of New York residents. Finally, a concluding remarks section closes the work. Proofs of the statements are deferred to the Appendix. 

\section{Literature Review} \label{sec:2}

\subsection{Disclosure Risk Limitation with categorical variables} \label{sub:DiscRisk}

In disclosure risk problems, we usually have microdata of $n$ individuals, where for each individual we can observe two distinct sets of variables: 1) some variables, usually called \textit{sensitive variables}, containing private information, e.g. health status or salary; 2) some identifying categorical variables, usually called \textit{key variables}, e.g. gender, age or job. Disclosure problems arise because an intruder may be able to identify individuals in the dataset by cross-classifying their corresponding key variables and matching them to some external source of information. If the matching is correct, the intruder will be able to disclose the information contained in the sensitive variables. 

Formally, let us assume we have $J$ categorical key variables in the dataset, observed for a sample of $n$ individuals, collected from a population of size $N$.  
Each variable has $n_{j}$ possible categories labelled, without loss of generality, from $1$ up to $n_j$. The observation for individual $i$, $X_{i}=(X_{i1},\ldots,X_{iJ})$, therefore takes values in the state space $\mathcal{C}:=\prod_{j=1}^{J}\{1\ldots,n_{j}\}$. 
This set has $K:=|\mathcal{C}|=\prod_{j=1}^{J}n_{j}$ values, corresponding to all possible cross-classification of the $J$ key variables. 
The information about the sample is usually given through the sample frequency vector $(f_{1},\ldots,f_{K})$, where $f_{i}$ counts how many individuals of the sample have been observed with the particular combination of cross-classified key variables corresponding to cell $i$. $(F_{1},\ldots,F_{K})$ denotes the corresponding vector frequencies when considering the whole population of $N$ individuals.


The earliest papers to consider disclosure risk problems include \cite{Bet90}, \cite{Dun86}, \cite{Dun89}, \cite{Lam93}. These works propose different measures of disclosure risk and possible ways to estimates them under different model choice.    
\cite{Ski02}, \cite{Ski94} review the most popular among measures of disclosure risk. These measures depend on the sample frequencies
$(f_{1},$ $\ldots,f_{K})$ and usually focus on small frequencies, especially those having frequency 1, called \emph{sample uniques}. The individuals belonging to these cells are those with the highest risk of their sensitive information being disclosed. Specifically, suppose that an individual is the only one both in the sample and in population to have a specific combination of key variables. Then, his key variables can be matched to an external database, and therefore this match will be perfect, i.e. correct with probability one, and his sensitive information will be therefore disclosed. 

We usually distinguish between two groups of \emph{measures of disclosure risk}: 
\begin{enumerate}
    \item \textbf{Record Level} (or per-record) measures: they assign a measure of risk for each data point. Among the most popular, there are
   \begin{equation} \label{riskindeces1}
   \begin{split}
       & r_{1k}=\mathbb{P}(F_{k}=1|f_{k}=1), \\
         &     r_{2k}=\mathbb{E}(1/F_{k}|f_{k}=1).
       \end{split}
   \end{equation}
   $k\in\{1,\ldots,K\}$.
   The first measure provides the probability that a sample unique is also population unique. The second tells the probability that if we select a sample unique and guess uniformly about his identity, we pick him correctly. The first measure is less conservative and is always smaller than the second. 
    \item \textbf{File level} measures: they provide an overall measure of risk for a dataset and are usually defined by aggregating the record level. Popular examples are
    \begin{equation} \label{riskindeces2}
        \tau_{1}=\sum_{k:f_{k}=1}r_{1k}, \ \ \
        \tau_{2}=\sum_{k:f_{k}=1}r_{2k}.
    \end{equation}
\end{enumerate}

These measures of disclosure risk are estimated using the data $(f_{1},\ldots,f_{K})$ under different modelling choices. For example, \cite{Ski08}, \cite{Shl10}  
consider the estimation of these measures under log-linear models for the population and sample frequencies. Under this model choice, the indexes \eqref{riskindeces1} and \eqref{riskindeces2}, can be derived in closed form and estimated using plug-in MLE estimators.
A different modelling approach, proposed in \cite{Man12},\cite{Man14}, is to apply grade of membership models, which provide very accurate estimates for \eqref{riskindeces2}. For a quite recent review on disclosure risk problems, the reader is referred to \cite{Mat11}. 

If the estimated values of \eqref{riskindeces1} and \eqref{riskindeces2} are too high, then the data curator should apply a disclosure limitation technique to the dataset before releasing it to the public. Some possibilities are for example  rounding, suppression of extreme values or entire variables, subsampling or perturbation techniques. See \cite{Wil01} for a review of different disclosure limitation techniques.

\subsection{Mutual Information}\label{sub:MutInf}

Let $X$ be a discrete random variable taking values on a finite set $\mathcal{X}$ and having probability mass function $p_{X}(x)$. The \emph{(Shannon) entropy} of $X$ is defined as
\begin{equation*} \label{entropy}
    H(X)=-\sum_{x\in \mathcal{X}}p_{X}(x)\log p_{X}(x)=-\mathbb{E}(\log(p_{X}(x)))
\end{equation*}
and it is a measure of uncertainty about the distribution of $X$. $H(X)$ is always non-negative, takes value $0$ when $p_{X}$ is a point mass in one of the support points and it is maximized when $p_{X}$ is uniform,  $p_{X}(x)=\frac{1}{|\mathcal{X}|}$ $\forall x\in\mathcal{X}$, in which case $H(X)=\log |\mathcal{X}|$.  \\
Similarly, given two discrete random variables $X$ and $Z$, their \emph{joint entropy} is defined as
\begin{equation*}
    H(X,Z)=-\sum_{x\in\mathcal{X}}\sum_{z\in\mathcal{Z}}p_{(X,Z)}(x,z)\log p_{(X,Z)}(x,z),
\end{equation*}
where $p_{(X,Z)}$ denotes the joint mass function on $\mathcal{X}\times\mathcal{Z}$. $ H(X,Z)$ measures the joint uncertainty of $X$ and $Z$ taken together. \\
Besides the \emph{conditional entropy} of $Z$ given $X$ is defined as
\begin{align}\label{cond.entr}
    H(Z|X) & = - \sum_{x,z} p_{X,Z}(x,z) \log(P_{Z|X=x}(z)) \\
    & = H(X,Z)-H(X) \nonumber
\end{align}
and quantifies the amount of information needed to describe the outcome of $Z$ given that the value of  $X$ is known. If $Z$ and $X$ are independent, the conditional entropy $H(Z|X)$ coincides with $H(Z)$. \\

The \textit{mutual information} between $X$ and $Z$ is defined as
\begin{equation*} \label{mutualinfo}
I(X,Z)=\sum_{z\in\mathcal{Z}}\sum_{x\in\mathcal{X}}P_{(X,Z)}(x,z)\log\left(\frac{p_{(X,Z)}(x,z)}{p_{X}(x)p_{Z}(z)}\right)
\end{equation*}
where $p_{X},p_{Z},p_{(X,Z)}$ are respectively the marginal and joint distributions of $X$ and $Z$. From the definition of $I(X,Z)$ it follows that
\begin{equation} \label{I-KL}
I(X,Z)=D_{KL}(p_{(X,Z)}||p_{X}p_{Z})
\end{equation}
where $D_{KL}$ denotes the Kullback-Leibler divergence. Therefore, $I(X,Z)$ measures the divergence between the joint distribution of $X$ and $Z$ and the product of their marginals. From \eqref{I-KL}, it also follows that $I(X,Z)\geq 0$, and $I(X,Z) = 0$ if and only if $X$ and $Z$ are independent.

An important equality connecting the mutual information $I(X,Z)$ with the marginal and joint entropies is
\begin{equation}\label{Inf-Ent}
    I(X,Z)=H(X)+H(Z)-H(X,Z).
\end{equation}
This formula is the base of the so-called $3H$ principle to estimate $I(X,Z)$, in which the three $H$ entropy terms on the right hand side are estimated from the data and plugged into \eqref{Inf-Ent} to obtain an estimate $\widehat{I(X,Z)}$.

For a review on entropy, mutual information and their properties, see for example \cite{Gib02,Gra11} and references therein.

\subsection{Differential Privacy} \label{sub:DiffPriv}

\textit{Differential Privacy} is a notion recently proposed in the computer science  literature by \cite{Dwo06,Dwo14} mathematically formalize the idea that the presence or absence of an individual in the raw data should have a limited impact on the transformed data, in order for the latter to be considered privatized.  Formally, let $X_{1:n}=(X_{1},\ldots,X_{n})$ be a set of observations, taking values in a state space $\mathcal{X}^{n}\subseteq \mathbb{R}^{n}$, containing sensitive information. A \textit{mechanism} is simply a conditional distribution $Q$ that, given the raw dataset $X_{1:n}$, returns a transformed dataset $Z_{1:k_{n}}=(Z_{1},\ldots,Z_{k_{n}})$, with $\mathcal{Z}^{k_n}\subseteq \mathbb{R}^{k_n}$, to be released to the public, where the sample sizes of $X_{1:n}$ and $Z_{1:n}$ are allowed to be different. Differential Privacy is a property of $Q$ that guarantees that it should be very difficult for an intruder to recover the sensitive information of $X_{1:n}$ by having access only to $Z_{1:k_{n}}$ and is defined as follows.

\begin{definition}[$\alpha$-Differential Privacy, \cite{Dwo06}] 
\label{Diff Priv}
The mechanism $Q$ satisfies $\alpha$-Differential Privacy if
\begin{equation} \label{diff priv}
\underset{S\in \sigma(\mathcal{Z}^{n})}{\sup}\frac{Q(Z_{1:n}\in S|X_{1:n})}{Q(Z_{1:n}\in S|X'_{1:n})} \leq \exp(\alpha)
\end{equation}  
for all $X_{1:n},X'_{1:n}\in \mathcal{X}^{n}$ s.t. $d_{H}(X_{1:n},X'_{1:n})=1$, where $d_{H}$ denotes the Hamming distance, $d_{H}(X_{1:n},X'_{1:n})=\sum_{i=1}^{n}\mathbb{I}(X_{i}\neq X'_{i})$ and $\mathbb{I}$ is the indicator function of the event inside brackets.
\end{definition}

For small values of $\alpha$ the right hand side of \eqref{diff priv} is approximately $1$. Therefore, if $Q$ satisfies Differential Privacy,  \eqref{diff priv} guarantees that the output database $Z_{1:n}$ has basically the same probability of having been generated from either one of two \emph{neighboring databases} $X_{1:n}$, $X'_{1:n}$, i.e. databases differing in only one entry. See \cite{Rin18} for a statistical viewpoint of differential privacy.

Differential Privacy has been studied in a wide range of problems, differing among them in the way data is collected and/or released to the end user. The two most important classifications are between Global vs Local privacy, and Interactive vs Non-Interactive models. In the \emph{Global (or Centralized) model} of privacy, each individual sends his data to the data curator who privatizes the entire data set centrally. Alternatively, in the \emph{Local (or Decentralized) model}, each user privatizes his own data before sending it to the data curator. In this latter model, data also remains secret to the possibly untrusted curator. In the \emph{Non-Interactive (or Off-line) model}, the transformed data set $Z_{1:n}$ is released in one spot and each end user has access to it to perform his statistical analysis. In the \emph{Interactive (or On-line) model} however, no data set is directly released to the public, but each end user can ask queries $f$ about $X_{1:n}$ to the data holder who will reply with a noisy version of the true answer $f(X_{1:n})$. 

 There have been many extensions and generalizations of the notion \eqref{diff priv} of Differential Privacy proposed over the last ten years, in order to accommodate for different areas of applications and state spaces of the input and output data. Among them, we mention $(\alpha,\delta)$-Differential Privacy (\cite{Dwo14}), vertex and edge Differential Privacy for network models (\cite{Bor15}), zero-mean Concentrated Differential Privacy (\cite{Bun16}), randomised differential privacy (\cite{Hal11}) or $\rho$ Differential Privacy (\cite{Cha13}, \cite{Dim17}), where the Hamming distance $d_{H}$ is \eqref{diff priv} is replaced by possibly any distance $\rho$, and many others.  However, since it is not possible to review all the many extensions of Differential Privacy here, we refer to \cite{Dwo14} for a quite updated review on different applications and extensions of Differential Privacy. 
To conclude this brief review, we recall one of the most important properties of any Differential Private mechanism: \emph{post processing}, see \cite{Dwo14}. This property guarantees that if the output $Z_{1:n}$ of any $\alpha$-Differential Private mechanism is further processed and gone through another mechanism (depending only on $Z_{1:n}$, and not on $X_{1:n}$), then the resulting output will also be  $\alpha$-Differential Private. Therefore, there will be no chance of any leak of privacy simply by post-processing the released data  $Z_{1:n}$.  

\section{An information-theoretic approach to PRAM using Differential Privacy}
\label{sec:3}

Post Randomization Method is a popular perturbation method for disclosure risk limitation. It is connected to randomized response techniques described by \cite{War65}. In the former approach, the raw data are perturbed by the data holder after having being collected, while in the latter, the perturbation is directly applied by the respondents during the interviewing process. We remind that PRAM was introduced by \cite{Koo97} and further explored by \cite{Gou97} and \cite{DeW97}. Given raw microdata, PRAM produces a new dataset where some of entries are randomly changed according to a prescribed probability mechanism. The randomness introduced by the mechanism implies that matching a record in the perturbed dataset may actually be a mismatch instead of a true match, hence making usual disclosure matching attempts less reliable.

\cite{Shl10} consider the problem of disclosure risk estimation when the microdata has gone through either a PRAM or data swapping process. They perturb the geographical key variable using a stochastic matrix $M$, i.e. every row of $M$ sums to one, where $M_{ij}$ provides the probability that an individual from location $i$ is changed to location $j$. \cite{Shl10} then proceed to discuss the problem of how to estimate the measures of risk presented in Subsection \ref{sub:DiscRisk}, but without providing any tangible rule on how to choose $M$, which is not the main goal of that paper. 

In this work, we propose a novel approach to choose the randomization matrix $M$ in PRAM. Specifically, we propose to choose it as the solution of a constrained maximization problem, in which  we maximize the mutual information between raw data $X_{1:n}$ and released data $Z_{1:n}$, under the constraint that the perturbation mechanism satisfies the Differential Privacy condition \eqref{diff priv}. Other optimization approaches for PRAM were already considered  by \cite{Wil99} and \cite{Wil00}, using different target functions and constraints. See also Section 5.5 of \cite{Wil01}. However, these choices usually result in a difficult maximization problems and often rely on approximation methods.

We argue that the choice of Mutual Information and Differential Privacy have several advantages. First, Mutual Information and Differential Privacy are very natural notions and popular measures of information similarity and privacy guaranty that have been widely considered in Information Theory and Machine Learning. Second, as it will be shown shortly, the resulting maximization problem reduces to a convex maximization problem under a set of linear constraints, hence it can be solved efficiently by well known optimization tools, like the Simplex method which is implemented in most of the commonly used computational softwares, like Matlab or R. Finally,  the level of privacy guaranteed  by the proposed methodology  is tuned by a single tuning parameter $\alpha$, which can be chosen by the data curator to achieve the desired level of privacy in a very simple manner. In subsection \ref{sub:Simulations}, we will show empirically how the choice of this parameter affects the estimation of the parameters, hence providing some evidence and guidance on how to choose it.

\subsection{Model of PRAM}

We propose to choose $M$ as the solution of the following constrained maximization program
    \begin{equation} \label{Maximization}
        \max_{M \ \text{satisfies} \ \eqref{diff priv}} I(X_{1:n},Z_{1:n}).
    \end{equation}
We will consider the case of randomly changing the values of a key variable with $S$ possible outcomes, e.g. the geographical location. $X_{i}\in \{1,\ldots,S\}$ is the corresponding categorical random variable, having probabilities $p=(p_{1},\ldots,p_{S})$, and therefore $\mathbb{P}(X_{i}=j)=p_{j}$. We consider the class of all randomizing matrices of the following form   
\begin{equation}\label{matrix_form}
M=
\begin{bmatrix}
    q_{1}       & \frac{1-q_{1}}{S-1} & \frac{1-q_{1}}{S-1} & \dots & \frac{1-q_{1}}{S-1} \\
    \frac{1-q_{2}}{S-1}       & q_{2} & \frac{1-q_{2}}{S-1} & \dots & \frac{1-q_{2}}{S-1} \\
    \hdotsfor{5} \\
    \frac{1-q_{S}}{S-1}       & \frac{1-q_{S}}{S-1} & \frac{1-q_{S}}{S-1} & \dots & q_{S}
\end{bmatrix}
\end{equation}
for an unknown parameter vector $q=(q_{1},\ldots,q_{S})$. This corresponds to the case in which, given that $X_{i}$ belongs to category $j$, then its transformed value $Z_{i}$ will either remain unchanged with probability $q_{j}$, or will be changed to one of the other $S-1$ categories, chosen uniformly at random, with probability $1-q_{j}$. 
Therefore, the  conditional distribution of $Z_{i}$ given $X_{i}$ is
\[
    Q(Z_{i}|X_{i})=q_{X_{i}}^{\mathbb{I}(Z_{i}=X_{i})}\left(\frac{1-q_{X_{i}}}{S-1}\right)^{\mathbb{I}(Z_{i}\not = X_{i})}.
\]
To underline the dependency on the vector $q$, we will sometimes write $Q_q$. It is easy to check that the marginal of $Z_{i}$ is given by
\begin{align}\label{marginal_Z}
  \mathbb{P}(Z_{i}=j)=   p_{j}q_{j}+\sum_{k\neq j}p_{k}\frac{1-q_{k}}{S-1} =: m_{j}
\end{align} 
for $j \in \{1,..,S \}$. We remark that the vector $m=(m_{1},\ldots,m_{S})$ can be computed in linear time in the dimension $S$ by first computing the quantity 
$ \sum_{k=1}^S p_{k}\frac{1-q_{k}}{S-1}. $\\

In the non interactive setting that we are considering, i.e. when $Z_i$ only depends on $X_i$, the conditional distribution of $Z_{1:n}$ factorizes and can be written as
\[
Q(Z_{1:n} | X_{1:n}) = \prod\limits_{i=1}^n Q(Z_i|X_i).
\]
Plugging it into \eqref{diff priv}, the Differential Privacy condition simplifies into 
\[
    \underset{Z_{i},X_{i}\not = X_{i}'}{\sup} \frac{Q(Z_{i}|X_{i})}{Q(Z_{i}|X_{i}')} \leq e^\alpha.
\]
Depending on the value of $Z_{i}$, the quotient \\
$Q(Z_{i}|X_{i})/Q(Z_{i}|X_{i}')$
 can take one of three values. If $Z_{i}=X_{i}$, then it is equal to $(S-1)q_{X_{i}}/(1-q_{X_{i}'})$. If $Z_{i}=X_{i}'$, then it is equal to  $(1-q_{X_{i}})/(S-1)q_{X_{i}'}$. Finally, if $Z_{i}$ is different from both $X_{i}$ and $X_{i}'$, then the quotient is equal to $ (1-q_{X_{i}})/(1-q_{X_{i}'})$. Therefore, the privacy condition specializes into the following set of constraints
\begin{equation} \label{max_constrained}
   \max \left( \frac{(S-1)q_{k}}{1-q_{k'}} , \frac{1-q_{k}}{(S-1)q_{k'}}, \frac{1-q_{k}}{1-q_{k'
}} \mathbb{I}(S \geq 3) \right) \leq e^\alpha
\end{equation}
for any couple $k \not= k' \in \{1,\ldots,S\}$. We notice that this set of conditions can be expressed as a linear constraint. Specifically,\vspace{0.5cm}

\textbf{Fact 1}: 
There exists a matrix $C$ and a vector  $b_\alpha$  (depending on $\alpha$) such that the set of differential privacy constraints \eqref{max_constrained} can be rewritten as the following linear constraint
\begin{equation}\label{linear_constraints}
    C q^T \leq b_\alpha,
\end{equation}
where $q$ is the vector $q=(q_1,..,q_S)$ and  $\leq$ denotes entry-wise inequality. $C$ and $b_\alpha$ are given in Appendix.\\
 
In general, computing $I(X,Z)$ takes of an order of $|\mathcal{X}||\mathcal{Z}|$ operations, meaning that here it should be quadratic in $S$. However, due to the particular form of the matrix $M$ considered here, this computation can be achieved linearly in $S$. Let us recall from Subsection \ref{sub:MutInf}, that $H(Z)$ denotes the Shannon entropy of the random variable $Z$ and $H(Z|X)$ the conditional entropy of $Z$ given $X$. To underline the dependency on $q$, we denote $f(q) := I(X,Z)$. We use the following known identity, which can immediately be derived from \eqref{Inf-Ent} and \eqref{cond.entr}, 
\[
 f(q) = I(X,Z) = H(Z) - H(Z|X),
\]
which leads to the simpler form
\begin{align*}
f(q) &= \sum_{x=1}^{S} p_x \left( q_x \log q_x + (1-q_x) \log \frac{1-q_x}{S-1}\right)\\
& \qquad\qquad\qquad- \sum_{z=1}^{S} m_z \log m_z
\end{align*}
where we recall that $m=(m_{1},\ldots,m_{J})$ denotes the marginal distribution of $Z$ given in \eqref{marginal_Z}. Let us start by noticing that $f$ is minimal, equal to $0$, for $q_1=..=q_S=\frac{1}{S}$. In the Appendix, we show that $f$ is convex in $q$, which, together with differential privacy constraint \eqref{linear_constraints}, implies that the problem \eqref{Maximization} is a linearly constrained convex program, i.e. we are maximizing a convex function under a set of linear constraints. As a consequence, the following proposition follows,

\begin{proposition}
\label{prop_optimal_max}
Any optimal $q$ solution of the program \eqref{Maximization}, lays within the vertices of the convex polytope formed by all the feasible points.
\end{proposition}

It follows from the previous proposition that finding the optimal matrix $M$ of general form \eqref{matrix_form} requires finding the vertices of the feasible set. In Section \ref{subsec:feasible} we will give some properties of this feasible set, which might make the search faster. In the following paragraph, we will provide the optimal $M$ for several sub-cases of \eqref{matrix_form}.

\subsection{Examples}
In this section we show how we can use Proposition \ref{prop_optimal_max} to give the explicit solutions of the program \eqref{Maximization} for several particular examples of interest.

\subsubsection{Binary key variable with symmetric $M$}
    We start from the simplest case of a categorical variable with only two possible categories denoted $\mathcal{X} = \{0,1\}$ and symmetric M with $q_1=q_2=q$. We will abuse our notations by writing $q$ both the scalar value in $[0,1]$ and the corresponding two-dimensional vector $(q,q)^T$ having both coordinates equal to this value. We are considering binary symmetric matrices of the following form,  
    \begin{equation*}
    M= \begin{bmatrix}
        q & 1-q  \\
        1-q &  q
        \end{bmatrix}.
    \end{equation*}
    In this setting, the Differential Privacy condition \eqref{max_constrained} specializes into 
\[
\max \left( \frac{q}{1-q}, \frac{1-q}{q} \right) \leq e^\alpha,
\]
    which simplifies to $q \in [\frac{1}{1+e^\alpha},\frac{e^\alpha}{1+e^\alpha}]$. In such a situation, the constrained maximization problem can actually be solved analytically by derivation of the target function. However, from Proposition \ref{prop_optimal_max}, it is already known that the optimal $q$ is among the boundaries of the feasible region. Let $\psi:\{0,1\}\rightarrow \{0,1\}$ be defined as $\psi(x) = 1-x$. Since $\psi$ is one-to-one, it follows that $I(X,Z)=I(X,\psi(Z))$. Moreover, by noticing that $\psi(Z)$ has conditional distribution $Q_{1-q}$, we can deduce that $I(X,\psi(Z))=f(1-q)$, and therefore $f(q)=f(1-q)$. Hence, the optimal $q$ are both boundaries points, $\frac{1}{1+e^\alpha}$ and $\frac{e^\alpha}{1+e^\alpha}$.
    
    There are two interesting properties appearing in this simple example. First, we understand that there are two solutions of the program. Second, these solutions are independent of $p$, the  marginal of $X$.

\subsubsection{Binary key variable with any $M$}
    The previous argument can be easily extended to the non-symmetric case,
     \begin{equation*}
    M= \begin{bmatrix}
        q_{1} & 1-q_{1}  \\
        1-q_{2} &  q_{2}
        \end{bmatrix}.
    \end{equation*}
    In this setting, the convex polytope generated by the linear constraints has four vertices, specifically $(q_1,q_2)$
belongs to the following set   
\[
\left\{ (1,0), (0,1), \Big(\frac{1}{1+e^\alpha},\frac{1}{1+e^\alpha}\Big), \Big(\frac{e^\alpha}{1+e^\alpha},\frac{e^\alpha}{1+e^\alpha}\Big)
\right\}
\] 
If either $(q_1,q_2)$ is equal to $(1,0)$ or $(0,1)$, then the Mutual Information $I(X_{1:n},Z_{1:n})$ is null, since $Z_{i}$ will be constant and independent of $X_{i}$. Therefore, the only optimal solutions are the two symmetric matrices derived in the symmetric case.

\subsubsection{Symmetric $M$}
    Let us now consider the case of a categorical variable with $S$ categories and symmetric $M$. Specifically, we consider $\mathcal{X} = \{1,\dots,S\}$ and $M$ of the form \eqref{matrix_form} with $q_1=q_2=\cdots=q_S$. We again 
    abuse our notation by denoting with $q$ both the scalar in $[0,1]$ and the corresponding $S$ dimensional vector with all entries equal to this value. The differential privacy condition \eqref{max_constrained} specializes into $\max \left( \frac{(S-1)q}{1-q}, \frac{1-q}{(S-1)q} \right) \leq e^\alpha$, which leads to $q \in [\frac{e^{-\alpha}}{S-1+e^{-\alpha}},\frac{e^\alpha}{S-1+e^\alpha}]$. As before, following from Proposition \ref{prop_optimal_max}, the optimal $q$ are the boundary values.

\subsection{Feasible set}\label{subsec:feasible}

In our experiments, we have experienced that routine optimization functions implemented in standard software, e.g. Matlab, can solve the optimization problem \eqref{Maximization} extremely quickly.  
However, when the number of possible categories $S$ becomes very large, the optimization might become time consuming. For this reason, in the following Proposition, we provide a description of all possible vectors $q=(q_{1},\ldots,q_{S})$ that can arise as vertices of the convex polytope generated by the Differential Privacy constraints \eqref{max_constrained} when $S$ is large enough. This result should help to speed up the search for the optimal vertices among all feasible points given by \eqref{max_constrained}.

\begin{proposition}\label{prop_feasible_set}
For $S \geq 4$, if $\alpha \leq \log (S+\sqrt{S(S-4)}-2)-\log 2$, then, up to permutations, the vertices of the convex polytope formed by all feasible points are:
\begin{enumerate}
    \item  $q_k =  v_\alpha$, $\forall k\in \{1,\ldots,S\}$;
    \item  $q_k  = v_{-\alpha}$, $\forall k\in \{1,\ldots,S\}$;
    \item $q_{k} = v_{i_k \alpha}$, with $i_k = \pm 1$ and $2 \leq  \#  \{k\ \text{s.t.} i_k = 1\} \leq S-2$,  
    $\forall k\in \{1,\ldots,S\}$;
    \item $q_1 = v_{\min}$, $q_{k } = v_\alpha$. , $\forall k\in \{2,\ldots,S\}$;
    \item $q_1 = v_{\max}$, $q_{k} = v_{-\alpha}$, $\forall k\in \{2,\ldots,S\}$;
\end{enumerate}
where $ v_x = \frac{e^{x}}{e^{x}+S-1}$, $v_{\min} = \frac{e^{-\alpha}}{e^{\alpha}+S-1}$ and $v_{\max} = \frac{e^{\alpha}}{e^{-\alpha}+S-1}$. 
\end{proposition}

Common values of $\alpha$ are generally within the range $[0,2]$, which means that the conditions of previous Proposition are satisfied when $S \geq 10$. Contrary to the symmetric case, the optimal $M$ will depend on $p$. In the following section, we will show some simulations illustrating for some values of $p$ which of the vertices in Proposition \ref{prop_feasible_set} are optimal.

\section{Simulations} \label{sec:4}

\subsection{Simulation Study} \label{sub:Simulations}

\begin{figure}[h!]
\centering
\subfigure[]{
\includegraphics[width=0.40\linewidth]{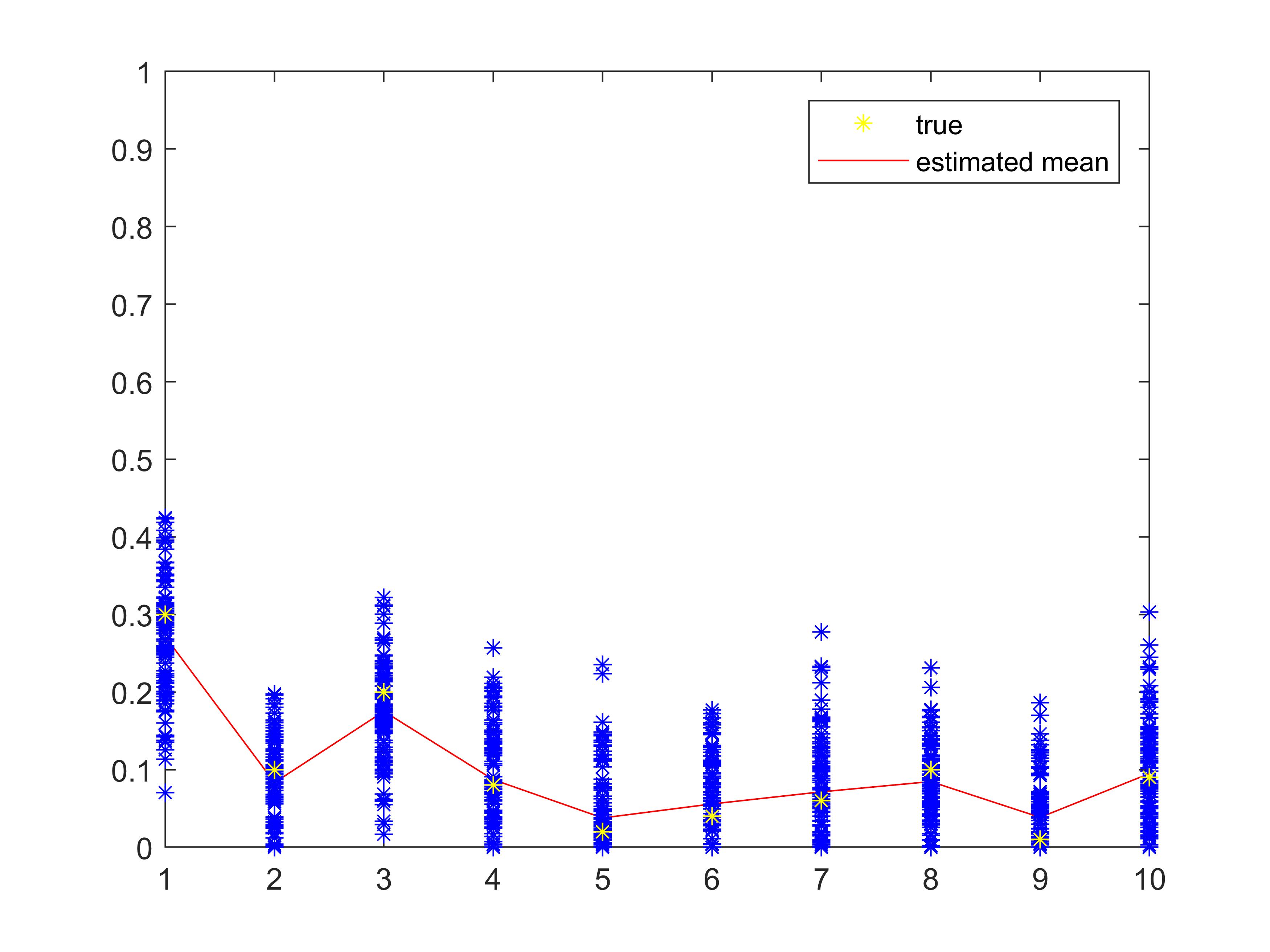}\label{alpha_05_n10e4_EX1}}
\subfigure[]{
\includegraphics[width=0.40\linewidth]{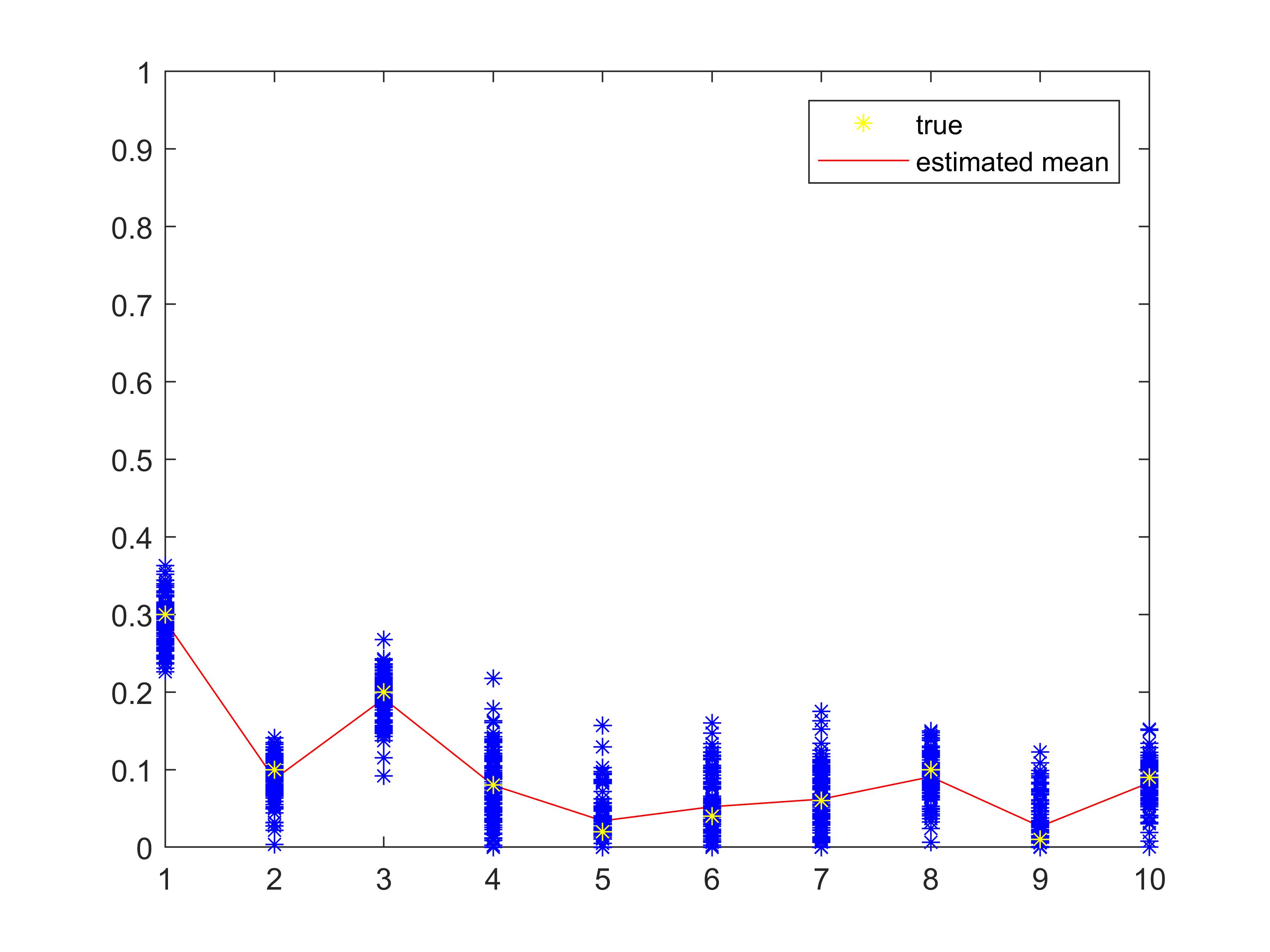}\label{alpha_10_n10e4_EX1}}\\
\subfigure[]{
\includegraphics[width=0.40\linewidth]{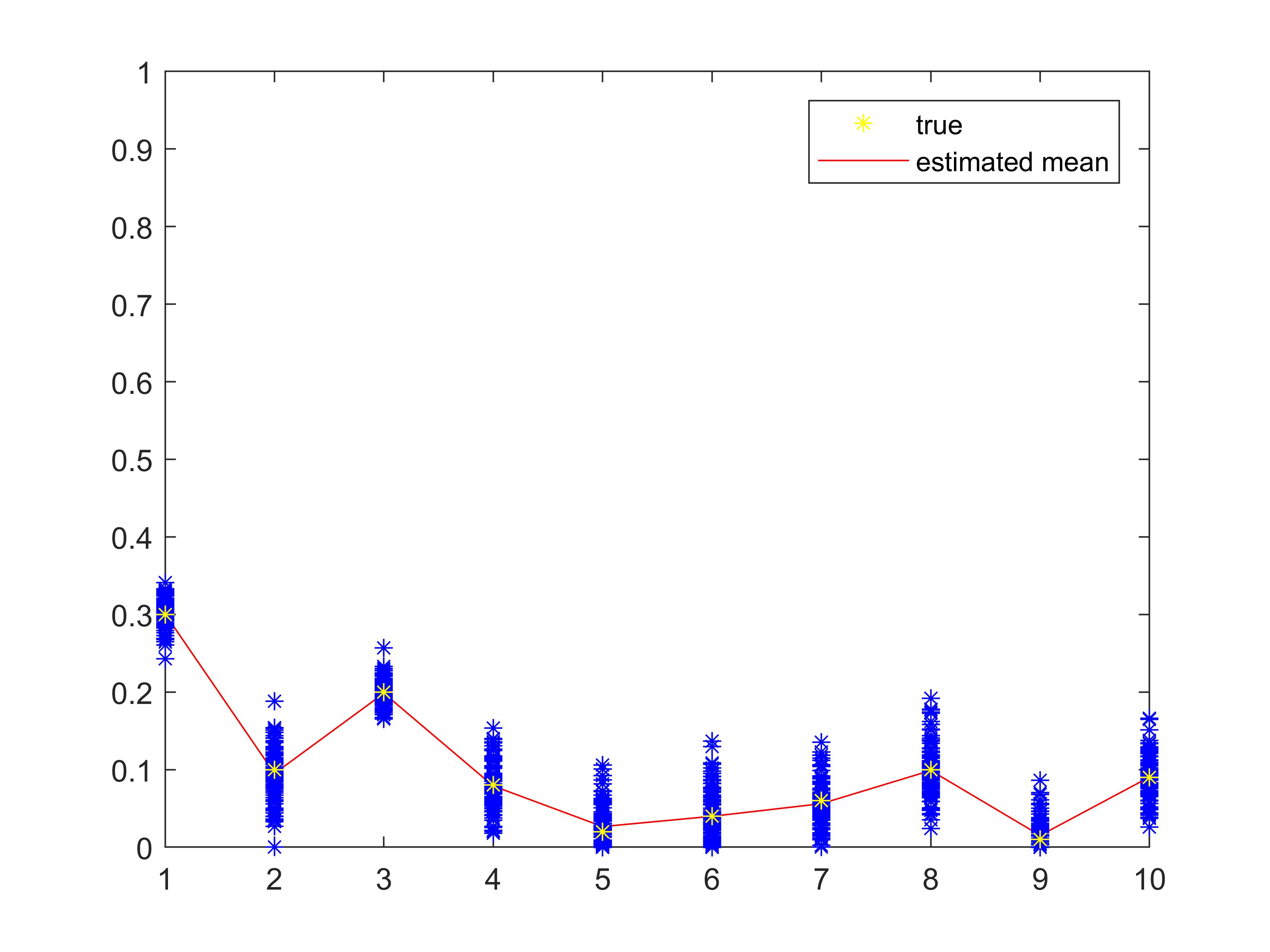}\label{alpha_15_n10e4_EX1}}
\subfigure[]{
\includegraphics[width=0.40\linewidth]{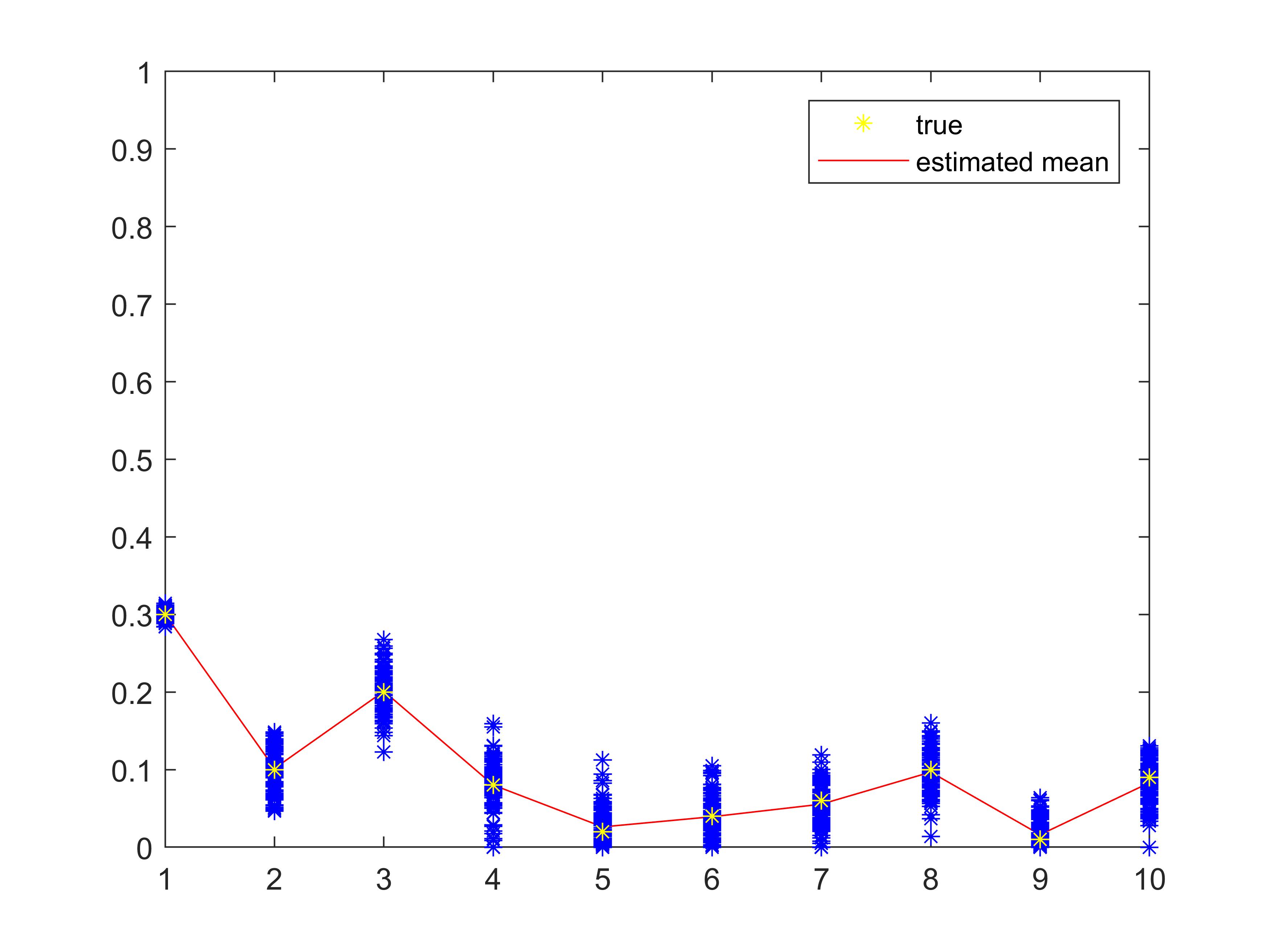}\label{alpha_20_n10e4_EX1}}\\
\caption{Scenario I: estimates of the true probabilities generating the data. The $x$-axis encodes the $S=10$ possible categories, for each one the yellow point represents the true probability $p_k$, while the solid red line connects the estimated probabilities averaged over $100$ iterations.}\label{fig:ex1}
\end{figure}

We consider different simulates scenarios, where the observations $X_{1:n}$ are generated from a categorical distribution with $S$ possible outcomes, having known probabilities $p=(p_1, \ldots , p_S)$.
In the first scenario, we set $S=10$ and consider the following vector of probabilities
\[
p=(0.3 ,0.1 ,0.2 ,0.08, 0.02 ,0.04, 0.06, 0.1, 0.01, 0.09).
\]
We consider the following values of $\alpha =0.5, 1, 1.5, 2$, and we determine the corresponding optimal vectors of $q= (q_1, \ldots , q_S)$ that solve 
\eqref{Maximization}. We select a sample size $n=10^4$. As explained in Section \ref{sec:3}, the Differential Privacy condition can be expressed as a set of linear constraints as in  \eqref{linear_constraints}. The optimal $q$ is then determined  numerically by solving the constrained maximization problem  via the optimization function in Matlab. 
Besides we have also generated the corresponding privatized dataset $Z_{1:n}$ using the determined values of $(q_1, \ldots , q_S)$ for the different choices of $\alpha$.
The determined values of $q$ are reported in Table \ref{tab:q}. From Proposition \ref{prop_feasible_set}, we know that, up to permutations, there are only $5$ possible different scenarios and the $q_k$'s may assume only 4 different values, corresponding to $v_\alpha, v_{-\alpha}, v_{{\rm min}}$ and $v_{{\rm max}}$. Hence in Table \ref{tab:q} we have reported the number of times the $q_k$'s assume these values for the different choices of $\alpha$.
\begin{table}[h!]
\centering
\begin{tabular}{|c||c|c|c|c|} \hline
$\alpha $ & $\# \, v_\alpha $  &  $\# \, v_{-\alpha}$ & $\# \, v_{\min}$ & $\# \,v_{\max}$ \\ \hline\hline
$0.5$ &   4 & 6 & 0 & 0 \\ \hline
$1$  &  5 & 5 & 0 & 0 \\ \hline
$1.5$ & 2 & 8 & 0 & 0 \\ \hline
$2$ & 0 & 9 & 0 & 1 \\ \hline 
		\end{tabular} 	
\caption{Scenario I: the number of times the $q_k$'s assume the four possible values $v_\alpha, v_{-\alpha}, v_{{\rm min}}$ and $v_{{\rm max}}$, under different choices of $\alpha$.}\label{tab:q}
\end{table}
In order to investigate the effect of differential privacy, for the four values of $\alpha$ considered here, we have reported the MLE of the vector of  probabilities $p$ obtained using the observed sample $Z_{1:n}$. The results are represented in Figure \ref{fig:ex1}, all the simulations are averaged over $100$ iterations. For each value of the categorical variable $ k \in \{1, \ldots , 10\}$, we have reported the estimated $p_k$'s, and each blue star corresponds to the MLE of $p_k$ in one of the $100$ experiments. The solid red line links the averaged estimates of the $p_k$'s over the $100$ runs, while the true values of the probabilities $p_k$ are represented in yellow.
It is apparent that as $\alpha$ increases, the estimates improve and the variability of the estimates decreases, hence the higher $\alpha$, the weaker the privacy mechanism.

\begin{center}
Figure \ref{fig:ex1} about here.
\end{center}

\begin{figure}[h!]
\centering
\subfigure[]{
\includegraphics[width=0.40\linewidth]{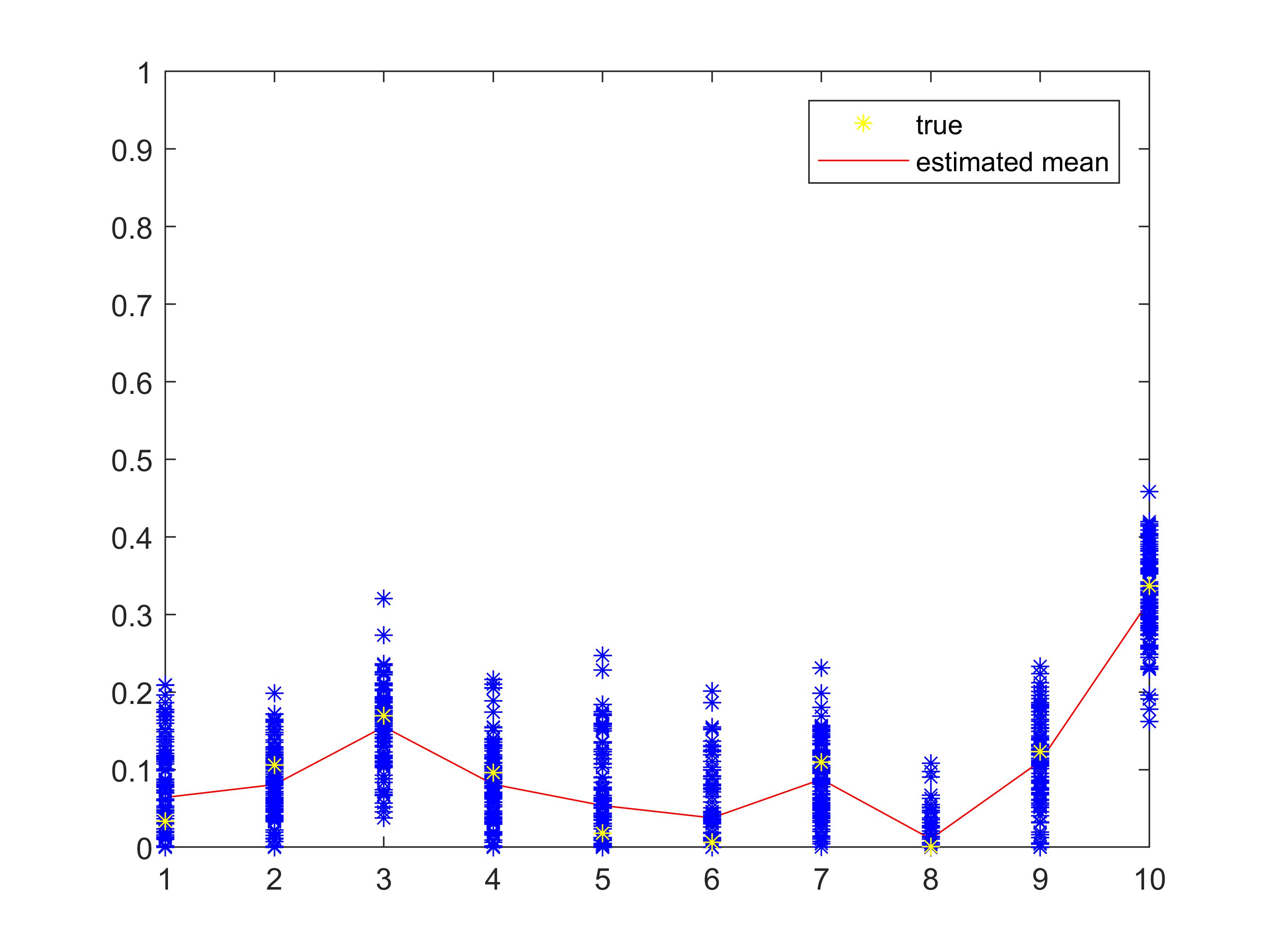}\label{alpha_05_n10e4_EX2}}
\subfigure[]{
\includegraphics[width=0.40\linewidth]{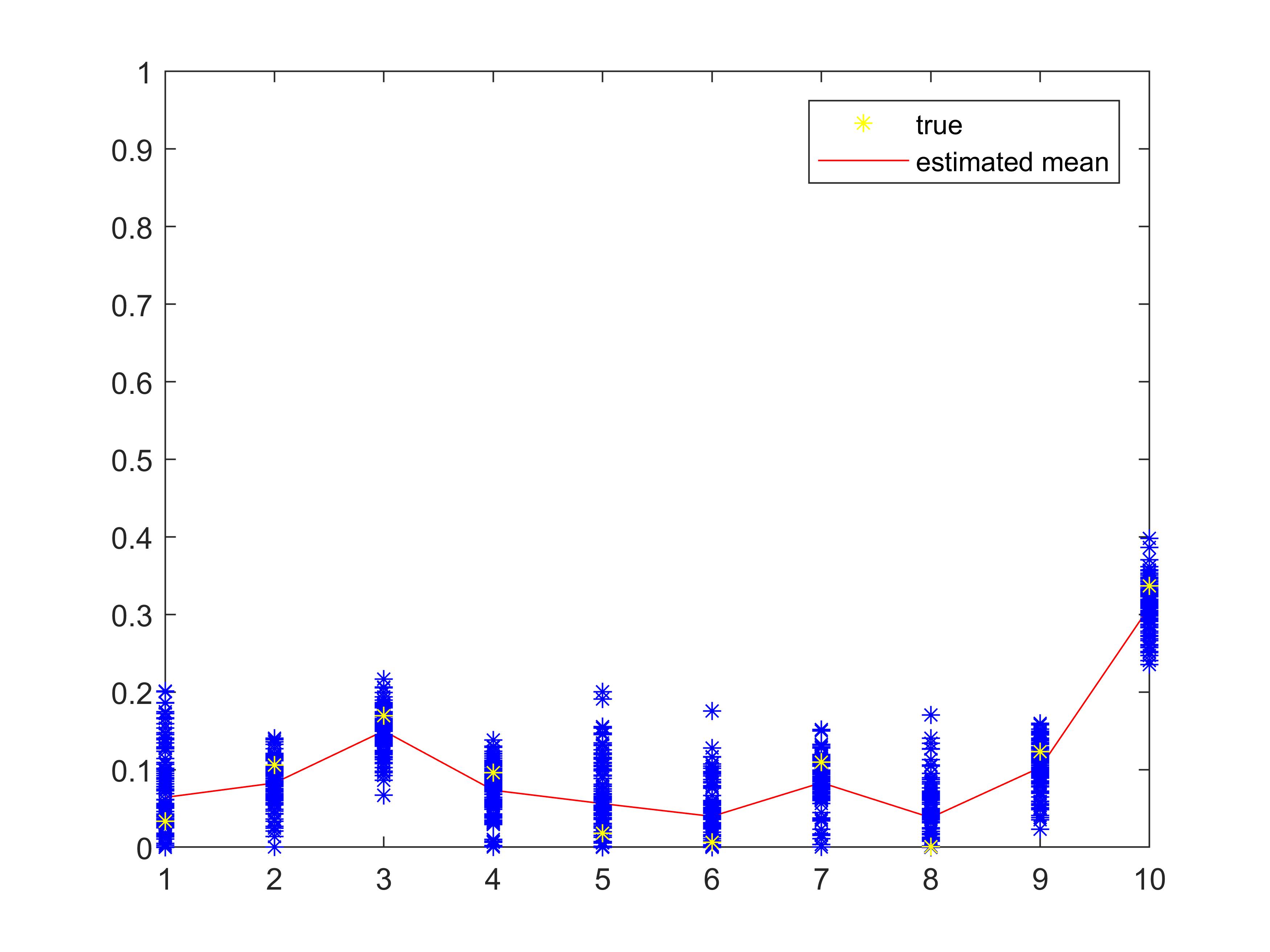}\label{alpha_10_n10e4_EX2}}\\
\subfigure[]{
\includegraphics[width=0.40\linewidth]{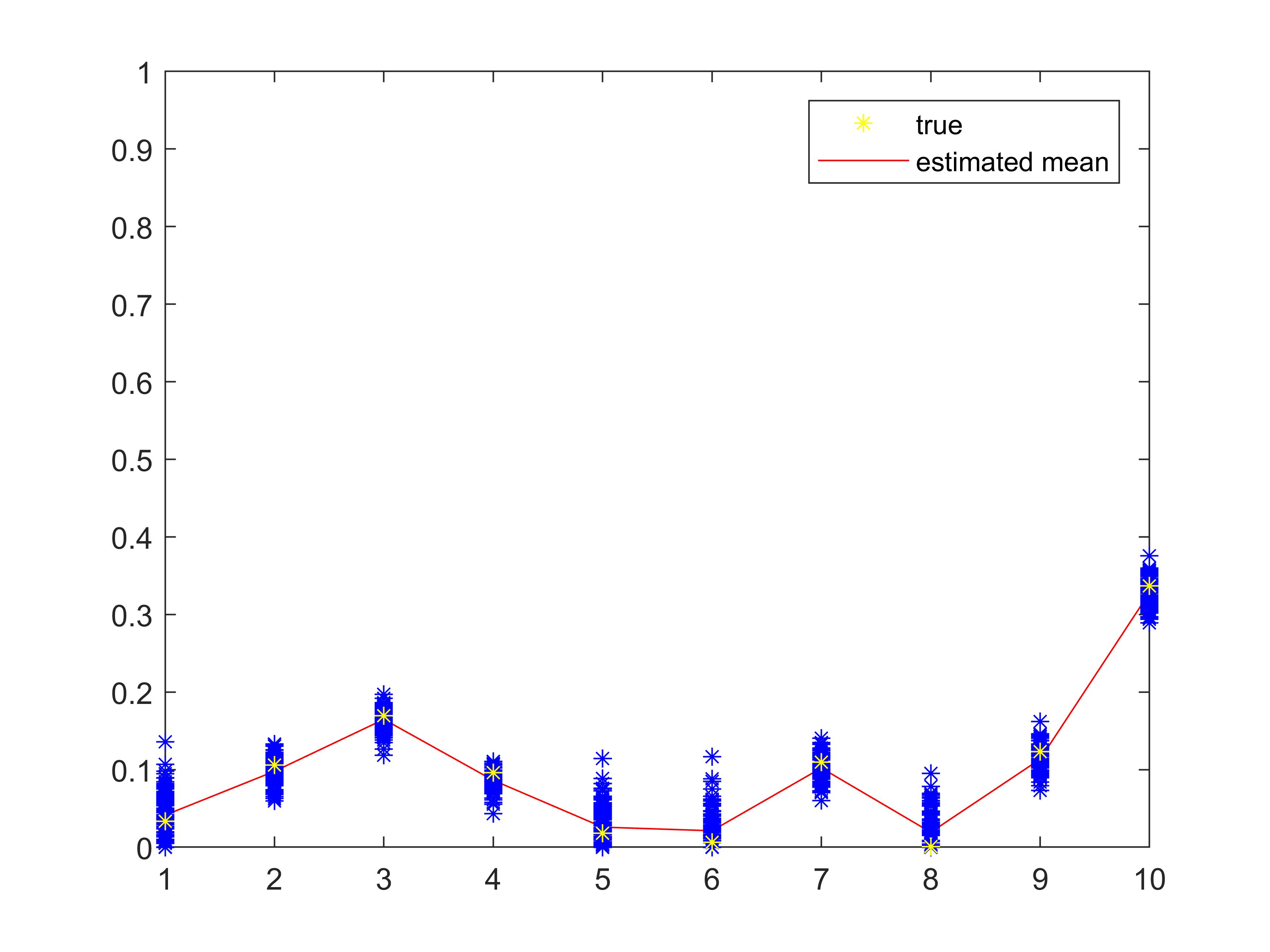}\label{alpha_15_n10e4_EX2}}
\subfigure[]{
\includegraphics[width=0.40\linewidth]{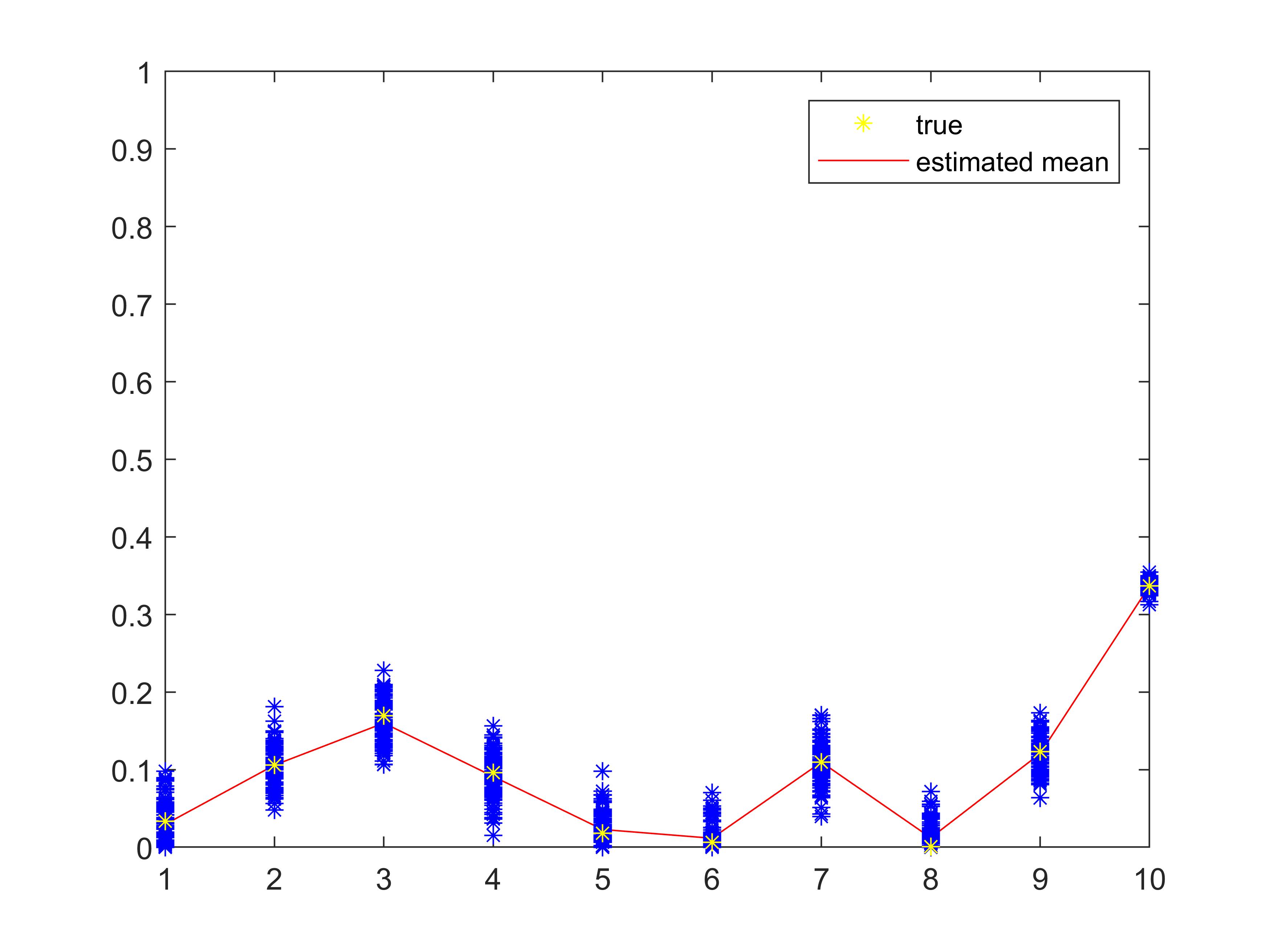}\label{alpha_20_n10e4_EX2}}\\
\caption{Scenario II: estimates of the true probabilities generating the data. The $x$-axis encodes the $S=10$ possible categories, for each one the yellow point represents the true probability $p_k$, while the solid red line connects the estimated probabilities averaged over $100$ iterations.}\label{fig:ex2}
\end{figure}

In the second scenario we have generated the data using the vector of probabilities
\begin{align*}
p &= (0.0336,  0.1059 ,   0.1697 ,   0.0962  ,  0.0180   , \\
 & \qquad \qquad 0.0062  ,  0.1097  ,  0.0005  ,  0.1233  ,  0.3369).
\end{align*}
As before we report the values of the $q_k$'s for different choices of $\alpha$ in Table \ref{tab:q2}, besides the estimated probabilities $p_k$'s are reported in Figure \ref{fig:ex2}. The simulations are averaged over $100$ iterations.

\begin{figure}[h!]
\centering
\subfigure[]{
\includegraphics[width=0.40\linewidth]{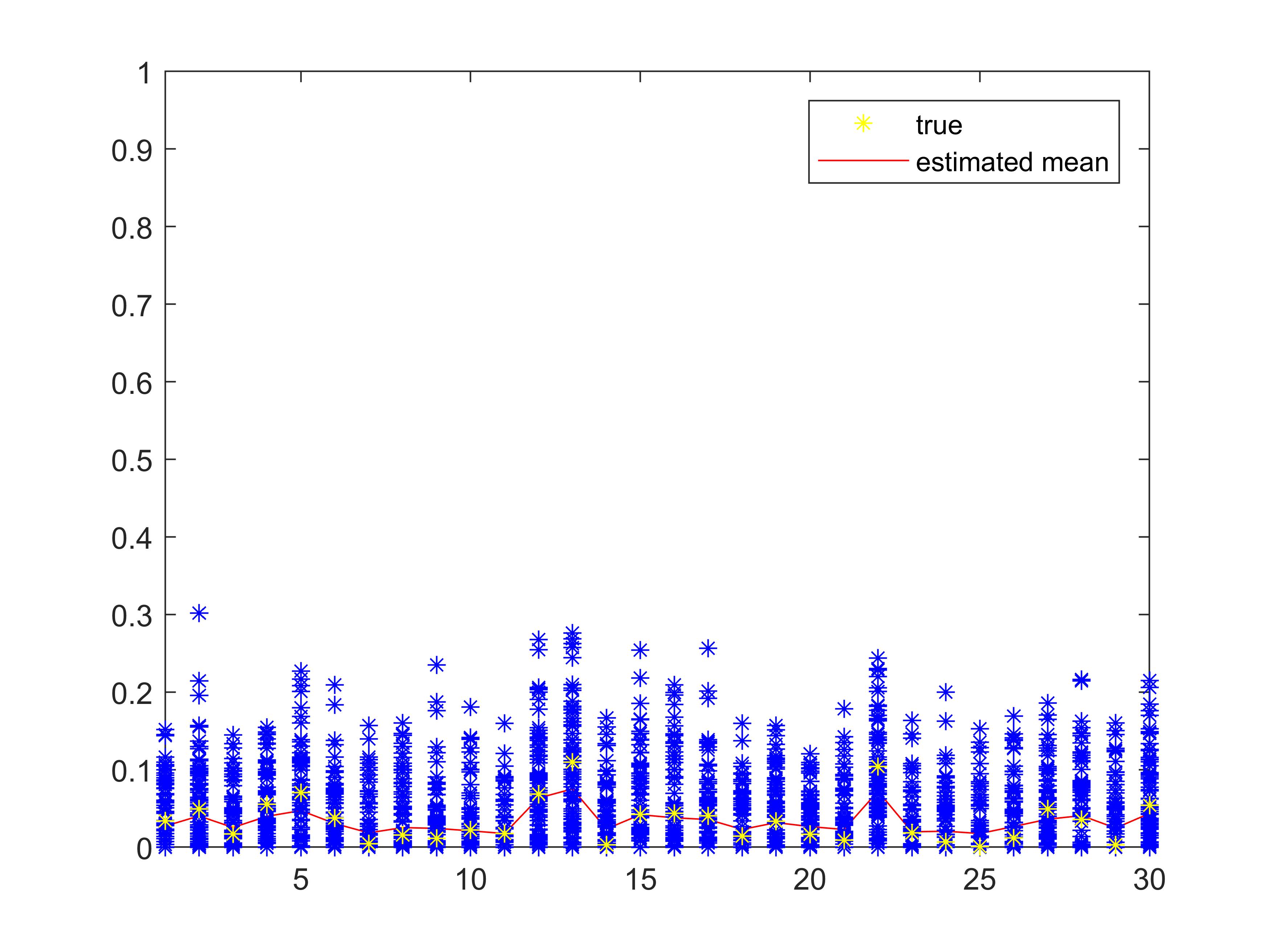}\label{alpha_05_n10e4_EX3}}
\subfigure[]{
\includegraphics[width=0.40\linewidth]{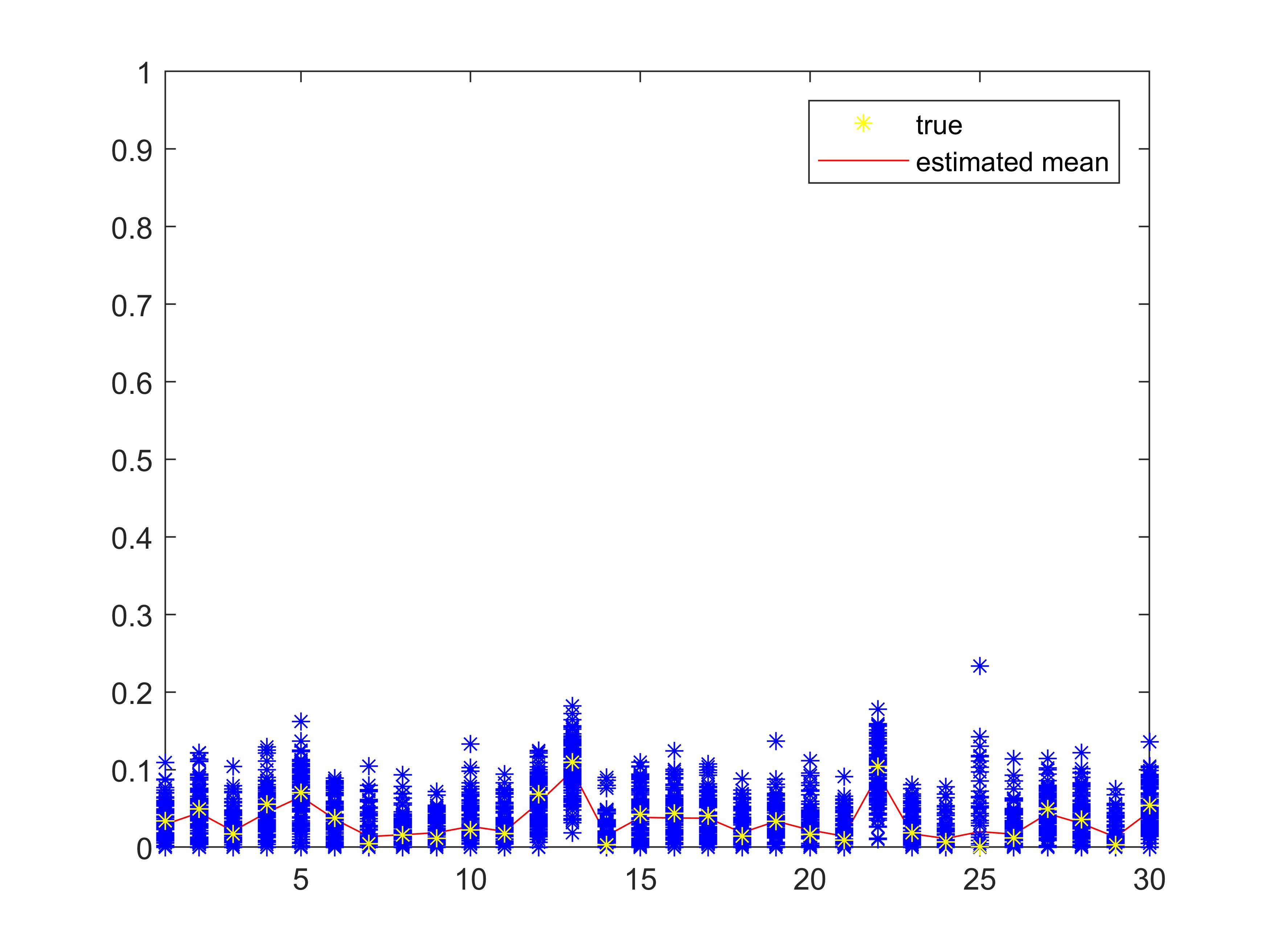}\label{alpha_10_n10e4_EX3}}\\
\subfigure[]{
\includegraphics[width=0.40\linewidth]{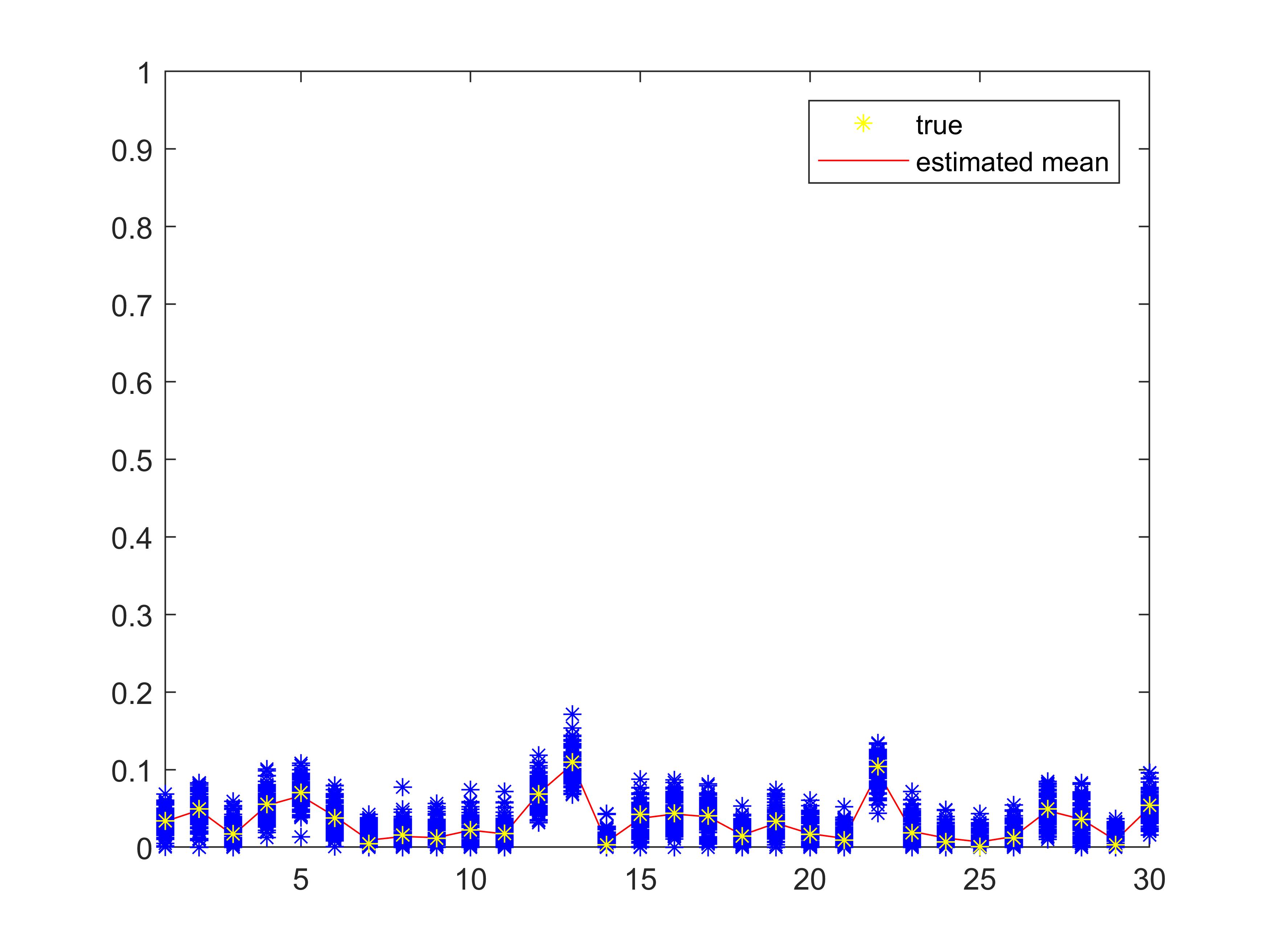}\label{alpha_15_n10e4_EX3}}
\subfigure[]{
\includegraphics[width=0.40\linewidth]{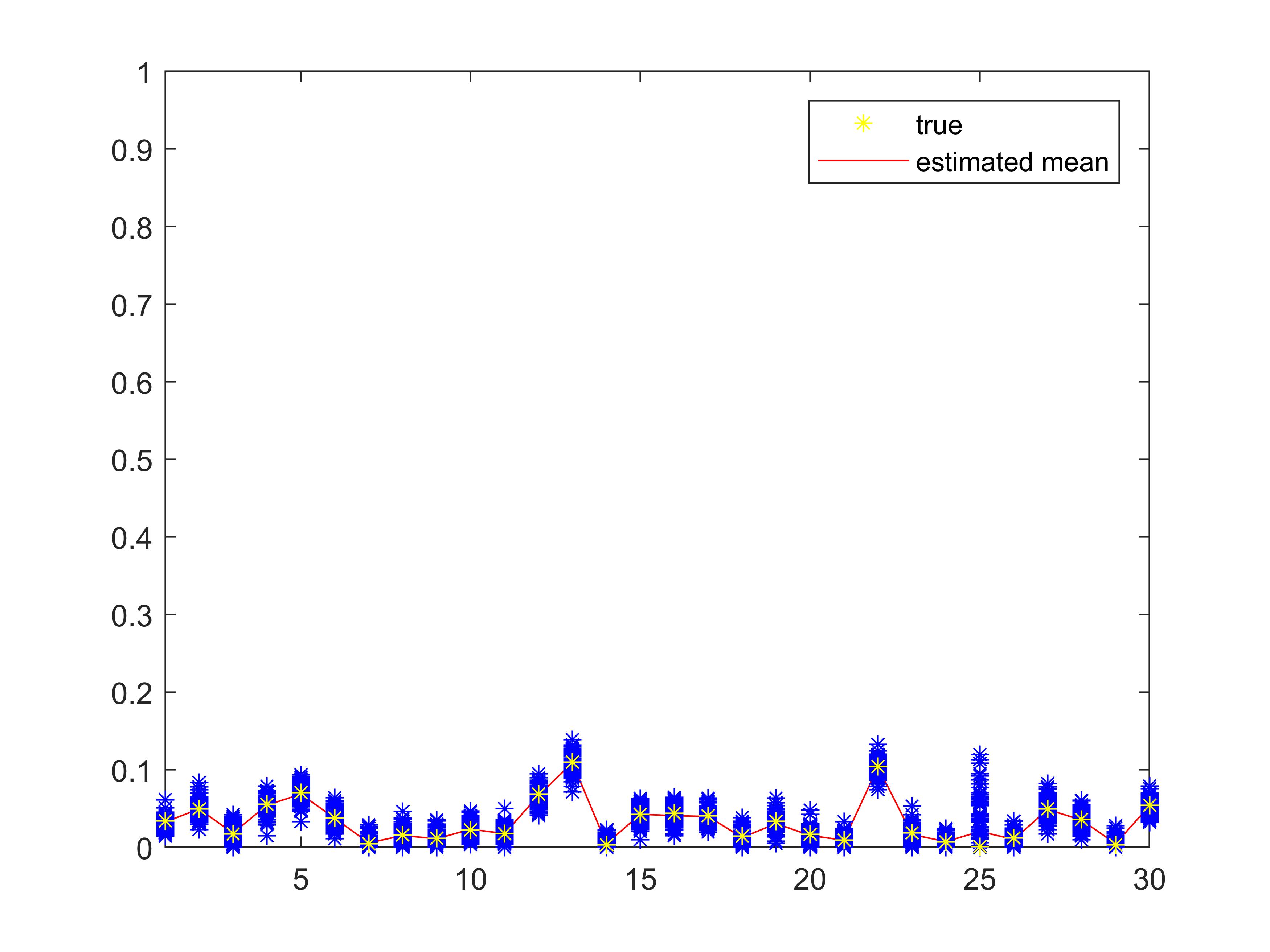}\label{alpha_20_n10e4_EX3}}\\
\caption{Scenario III: estimates of the true probabilities generating the data. The $x$-axis encodes the $S=30$ possible categories, for each one the yellow point represents the true probability $p_k$, while the solid red line connects the estimated probabilities averaged over $100$ iterations.}\label{fig:ex3}
\end{figure}

\begin{table}[h!]
\centering
\begin{tabular}{|c||c|c|c|c|} \hline
$\alpha $ & $\# \, v_\alpha $  &  $\# \, v_{-\alpha}$ & $\# \, v_{\min}$ & $\# \,v_{\max}$ \\ \hline\hline
$0.5$ &   7 & 3 & 0 & 0 \\ \hline
$1$  &  6 & 4 & 0 & 0 \\ \hline
$1.5$ & 6 & 4 & 0 & 0 \\ \hline
$2$ & 0 & 9 & 0 & 1 \\ \hline 
		\end{tabular} 	
\caption{Scenario II: the number of times the $q_k$'s assume the four possible values $v_\alpha, v_{-\alpha}, v_{{\rm min}}$ and $v_{{\rm max}}$, under different choices of $\alpha$.}\label{tab:q2}
\end{table}
\begin{center}
Figure \ref{fig:ex2} about here.
\end{center}
We consider now a third scenario, in which $S=30$ and  we have generated the data using the vector of probabilities $p$ obtained as a normalization of $30$ independent gamma random variables with parameters $(1,5)$, more precisely we have generated $G_k \sim  {\rm Gamma}(1,5)$ for $k=1, \ldots , 30$ and we have put $p_k := G_k/ \sum_{s=1}^S G_s$.
As before we report the vectors of $q$ for different values of $\alpha$ in Table \ref{tab:q3} and the estimated probabilities averaged over $100$ iterations in Figure \ref{fig:ex3}, where again $n=10^4$ is the sample size.

\begin{figure}[h!]
\centering
\subfigure[]{
\includegraphics[width=0.40\linewidth]{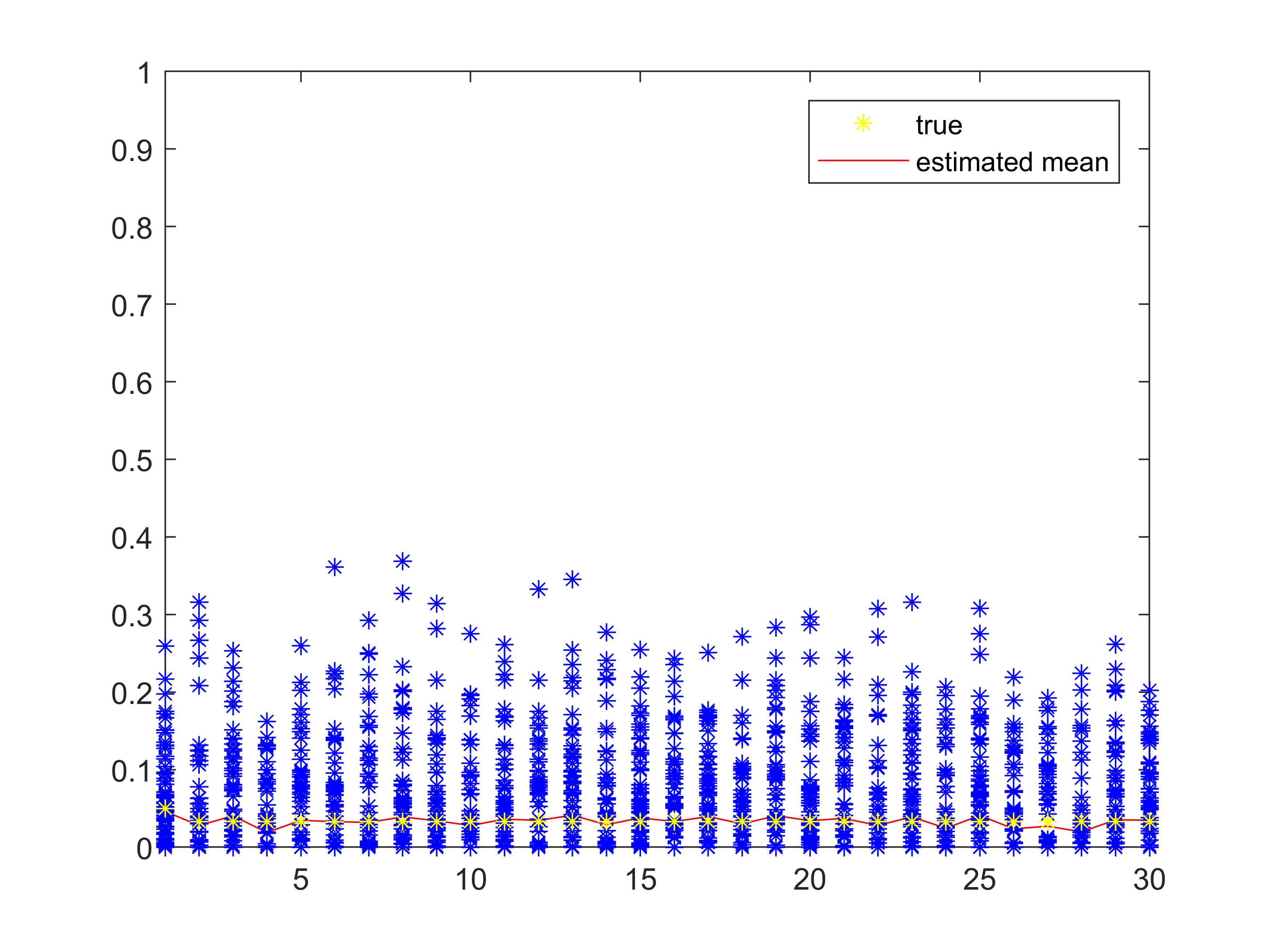}\label{alpha_05_n10e4_EX4}}
\subfigure[]{
\includegraphics[width=0.40\linewidth]{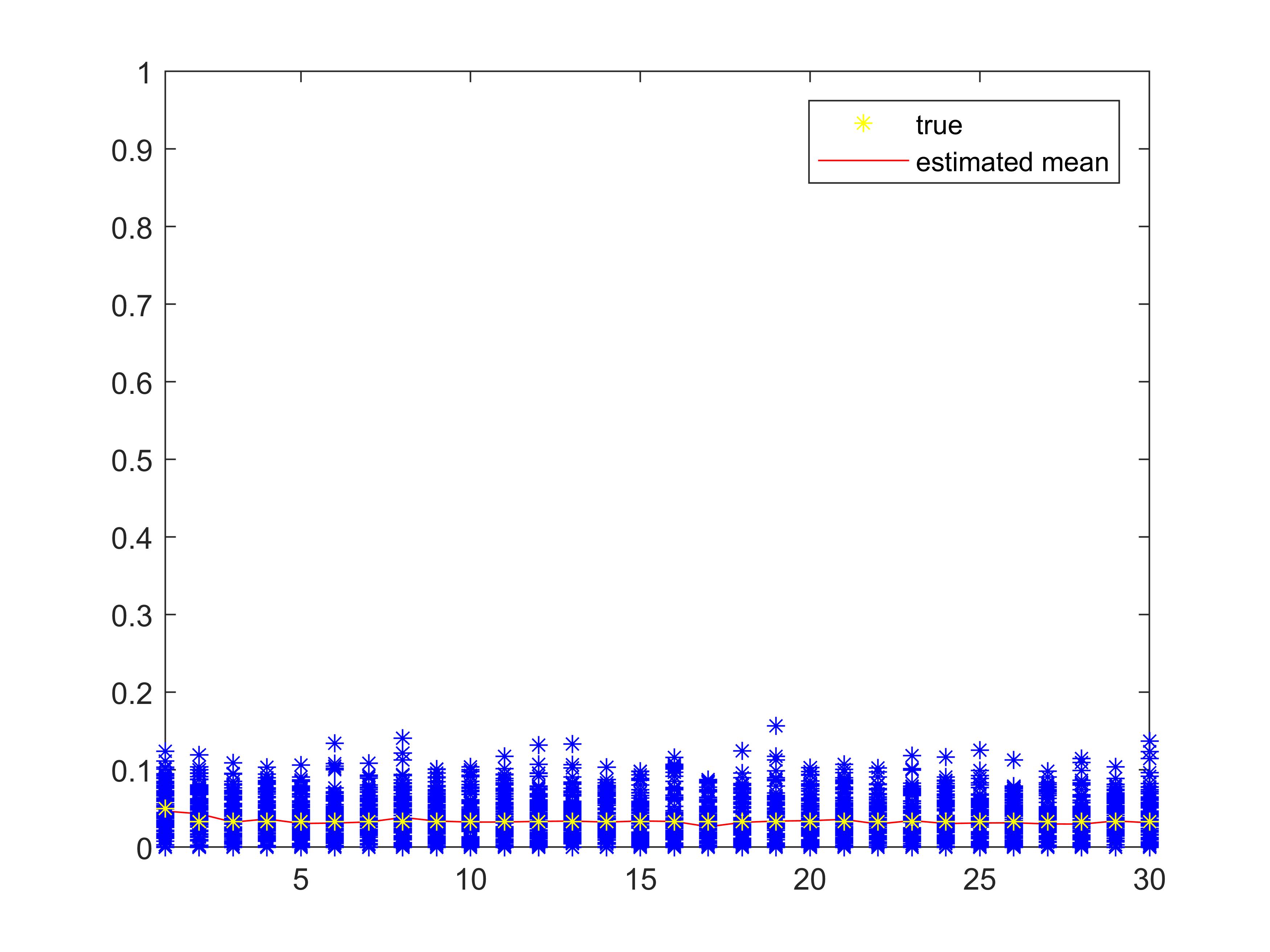}\label{alpha_10_n10e4_EX4}}\\
\subfigure[]{
\includegraphics[width=0.40\linewidth]{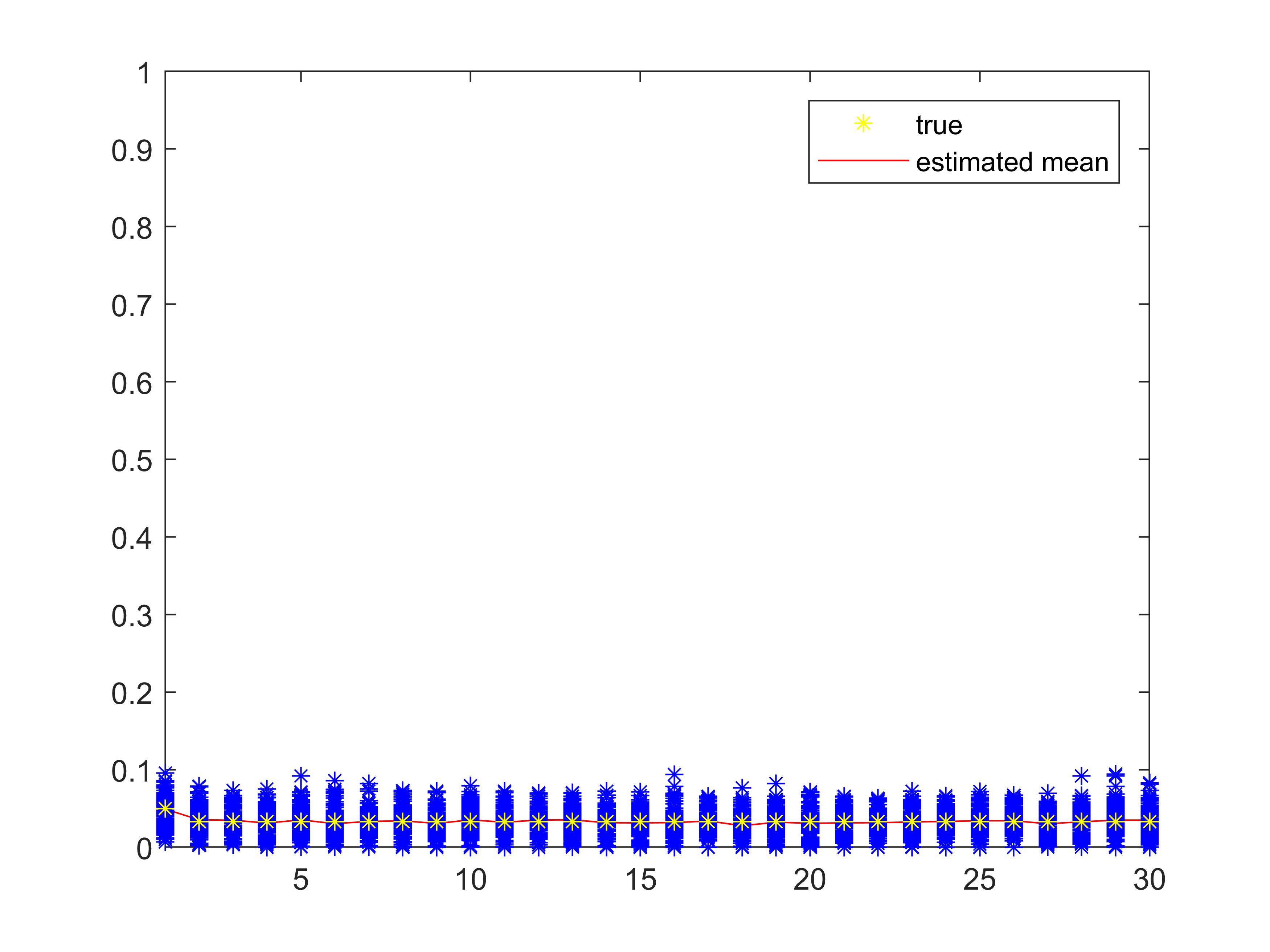}\label{alpha_15_n10e4_EX4}}
\subfigure[]{
\includegraphics[width=0.40\linewidth]{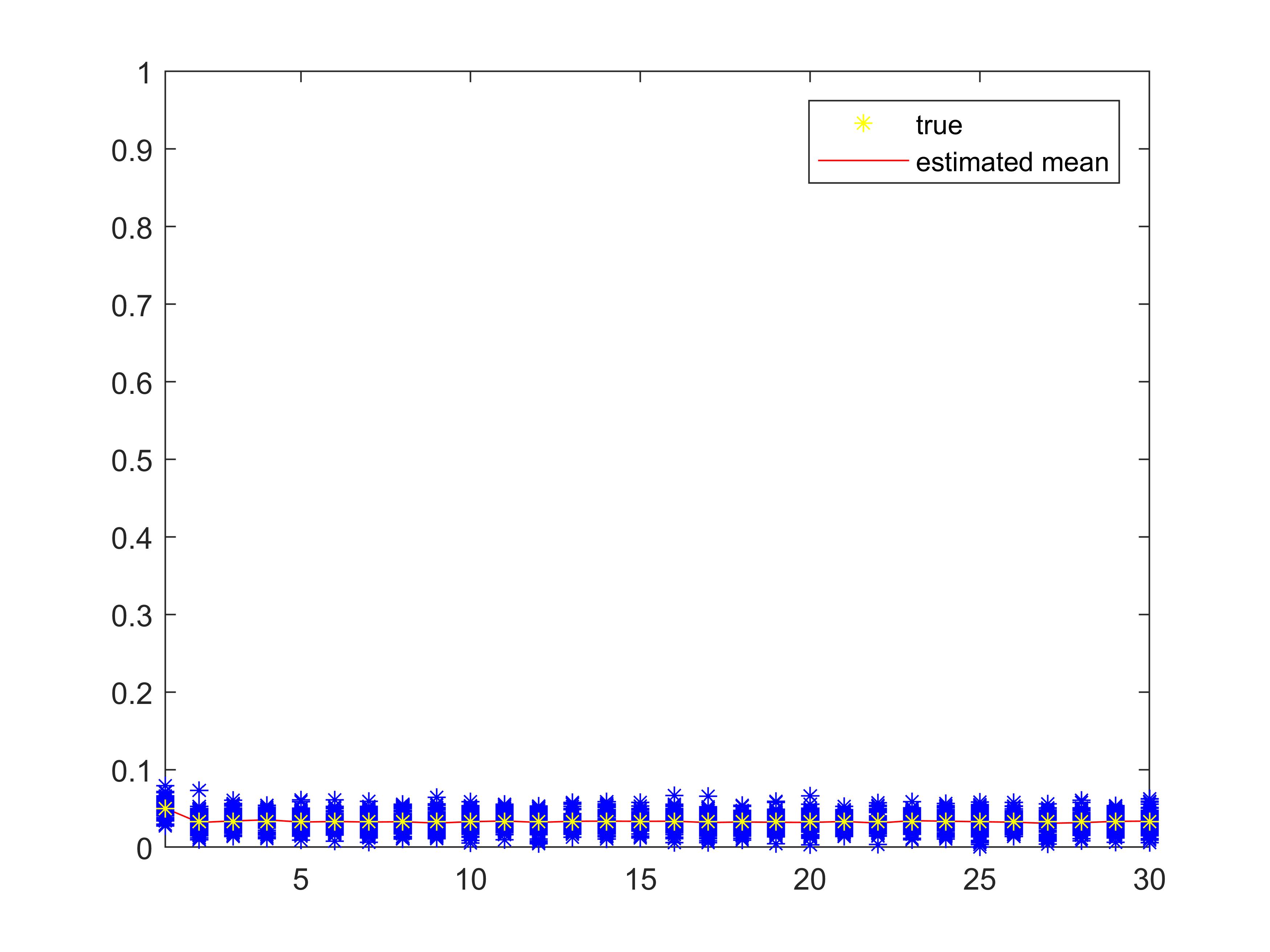}\label{alpha_20_n10e4_EX4}}\\
\caption{Scenario IV: estimates of the true probabilities generating the data. The $x$-axis encodes the $S=30$ possible categories, for each one the yellow point represents the true probability $p_k$, while the solid red line connects the estimated probabilities averaged over $100$ iterations.}\label{fig:ex4}
\end{figure}

\begin{table}[h!]
\centering
\begin{tabular}{|c||c|c|c|c|} \hline
$\alpha $ & $\# \, v_\alpha $  &  $\# \, v_{-\alpha}$ & $\# \, v_{\min}$ & $\# \,v_{\max}$ \\ \hline\hline
$0.5$ &   29 & 0 & 1 & 0 \\ \hline
$1$  &  29 & 0 & 1 & 0 \\ \hline
$1.5$ & 30 & 0 & 0 & 0 \\ \hline
$2$ & 29 & 0 & 1 & 0 \\ \hline 
		\end{tabular} 
\caption{Scenario III: the number of times the $q_k$'s assume the four possible values $v_\alpha, v_{-\alpha}, v_{{\rm min}}$ and $v_{{\rm max}}$, under different choices of $\alpha$.}	\label{tab:q3}
\end{table}
\begin{center}
Figure \ref{fig:ex3} about here.
\end{center}
In the last scenario IV, we assume again that $S=30$ and  we have generated the data using the vector of probabilities $p$ defined by
\[
p_1=0.05 , \quad  p_k = 0.95/29 \, \text{ for } k \geq 2. 
\]
We report the vectors of $q$ for different values of $\alpha$ in Table \ref{tab:q4} and the estimated probabilities averaged over $100$ iterations in Figure \ref{fig:ex4}, where the sample size equals $n=10^4$.
\begin{table}[h!]
\centering
\begin{tabular}{|c||c|c|c|c|} \hline
$\alpha $ &   $\# \, v_\alpha $  &  $\# \, v_{-\alpha}$ & $\# \, v_{\min}$ & $\# \,v_{\max}$ \\ \hline\hline
$0.5$ &   0 & 29 & 0 & 1 \\ \hline
$1$  &  30 & 0 & 0 & 0 \\ \hline
$1.5$ & 30 & 0 & 0 & 0 \\ \hline
$2$ & 30 & 0 & 0 & 0 \\ \hline 
		\end{tabular} 
\caption{Scenario IV: the number of times the $q_k$'s assume the four possible values $v_\alpha, v_{-\alpha}, v_{{\rm min}}$ and $v_{{\rm max}}$, under different choices of $\alpha$.}	\label{tab:q4}
\end{table}
\begin{center}
Figure \ref{fig:ex4} about here.
\end{center}

\subsection{Real Data}
\label{sub:RealData}

\begin{figure}[h!]
\centering
\subfigure[]{
\includegraphics[width=0.40\linewidth]{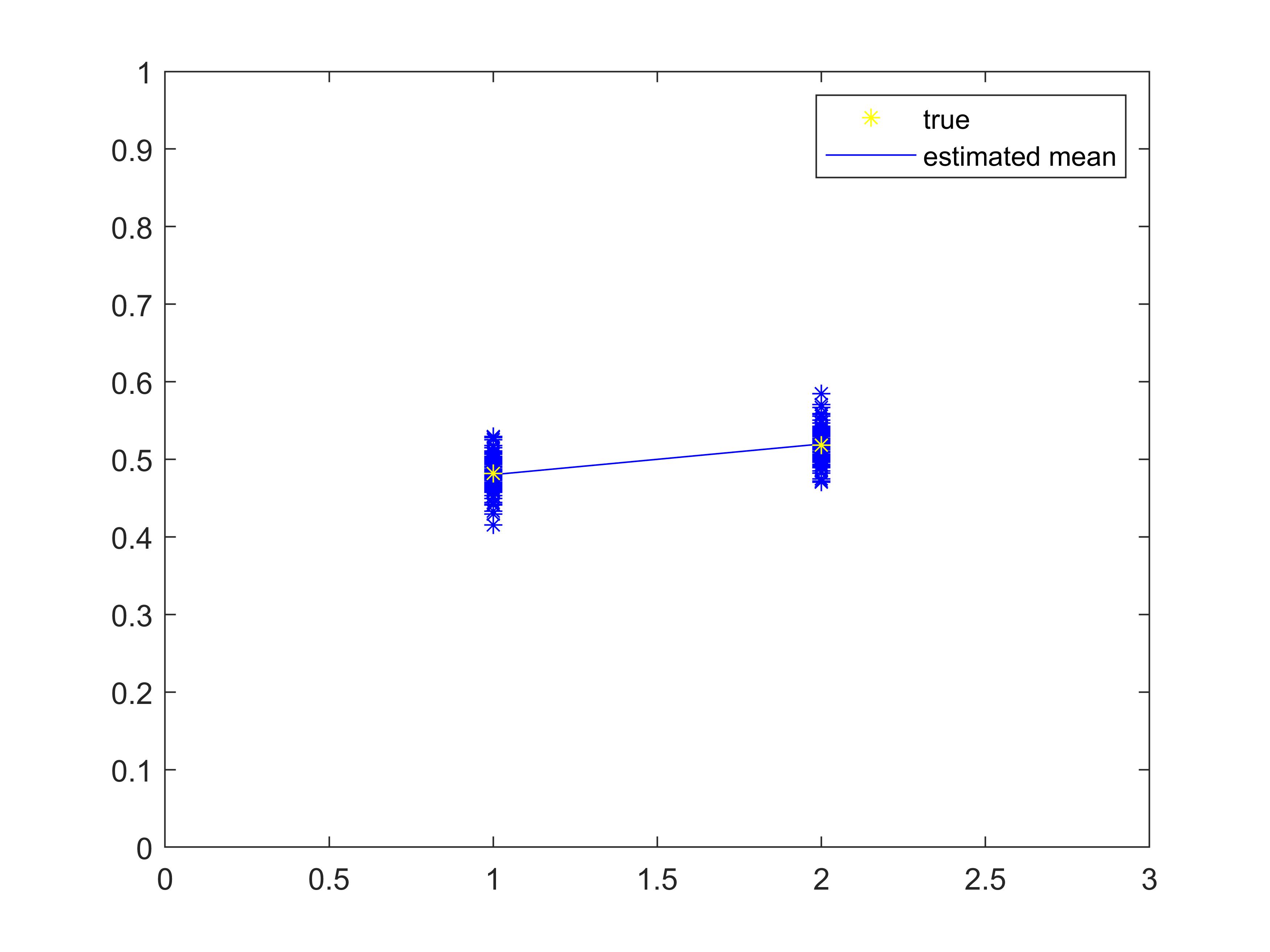}}
\subfigure[]{
\includegraphics[width=0.40\linewidth]{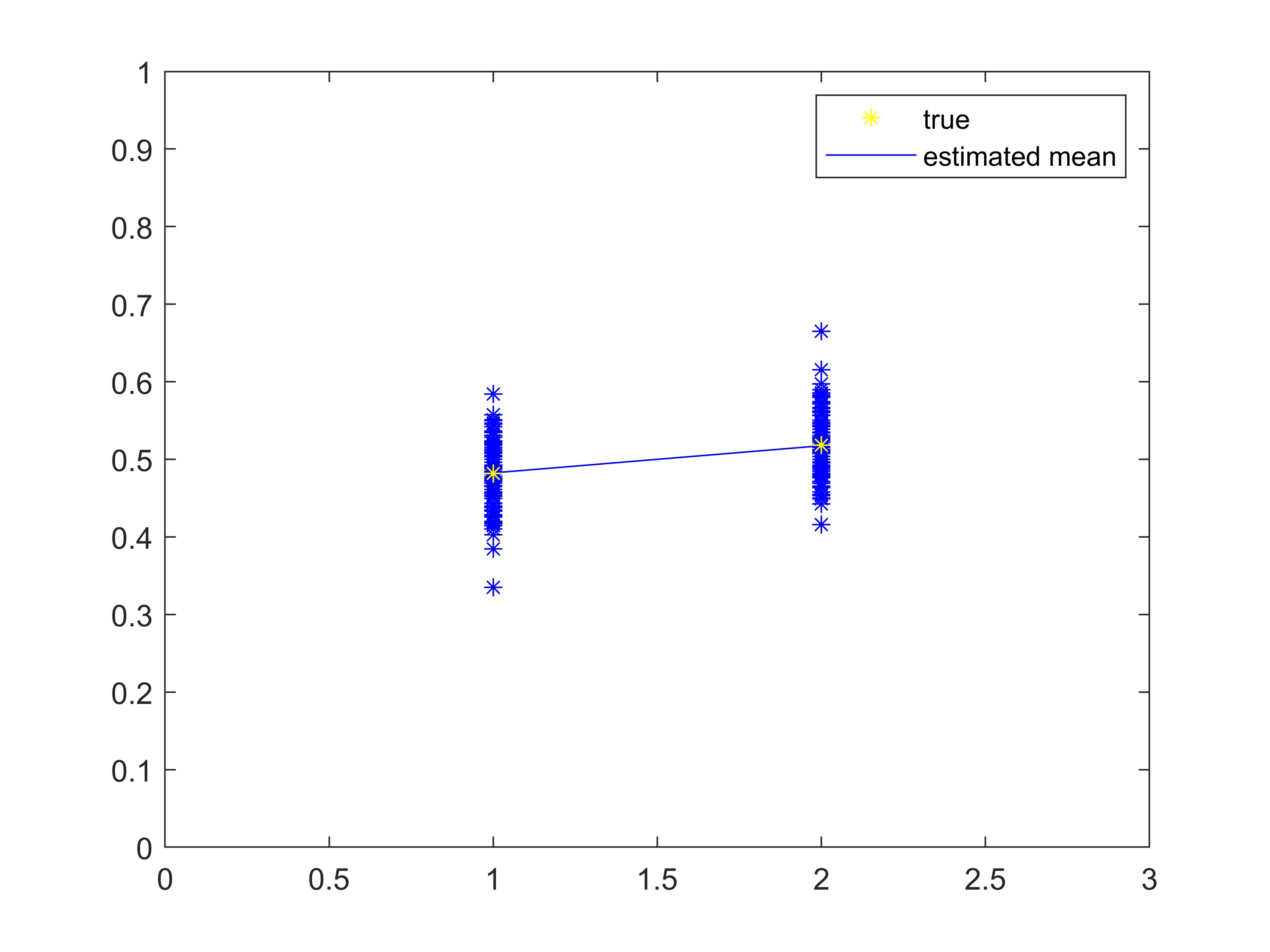}}\\
\subfigure[]{
\includegraphics[width=0.40\linewidth]{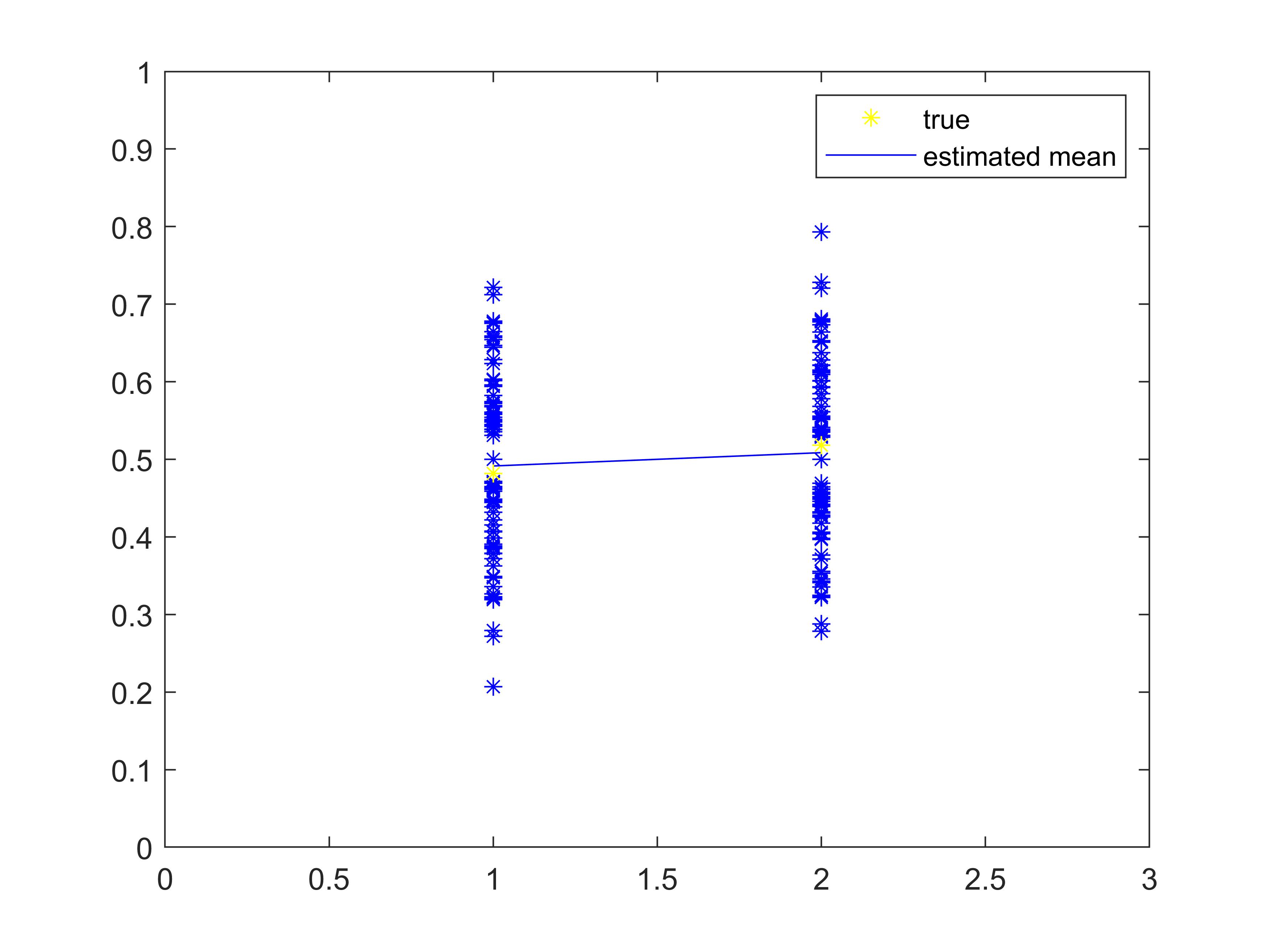}}
\subfigure[]{
\includegraphics[width=0.40\linewidth]{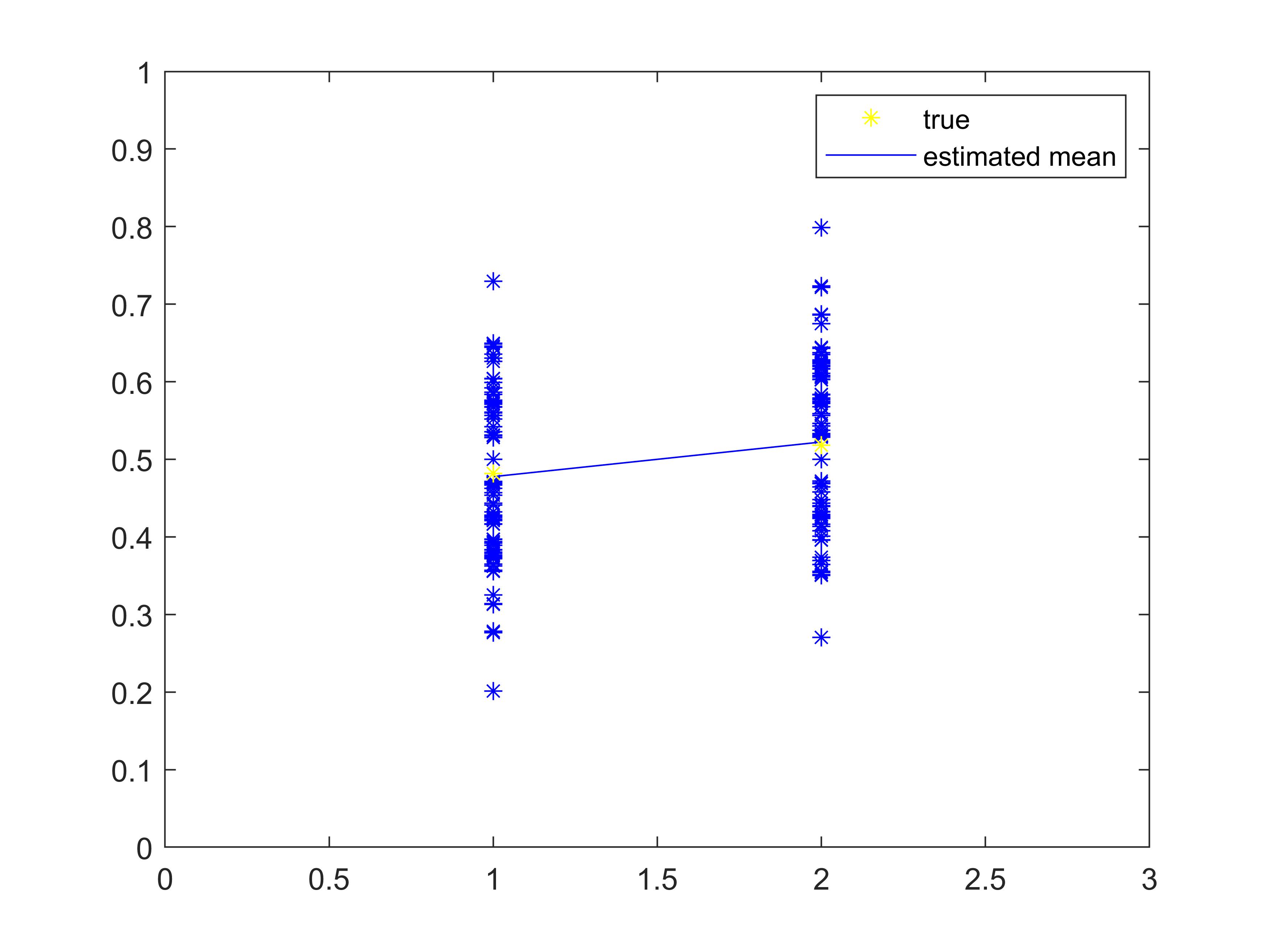}}\\
\subfigure[]{
\includegraphics[width=0.40\linewidth]{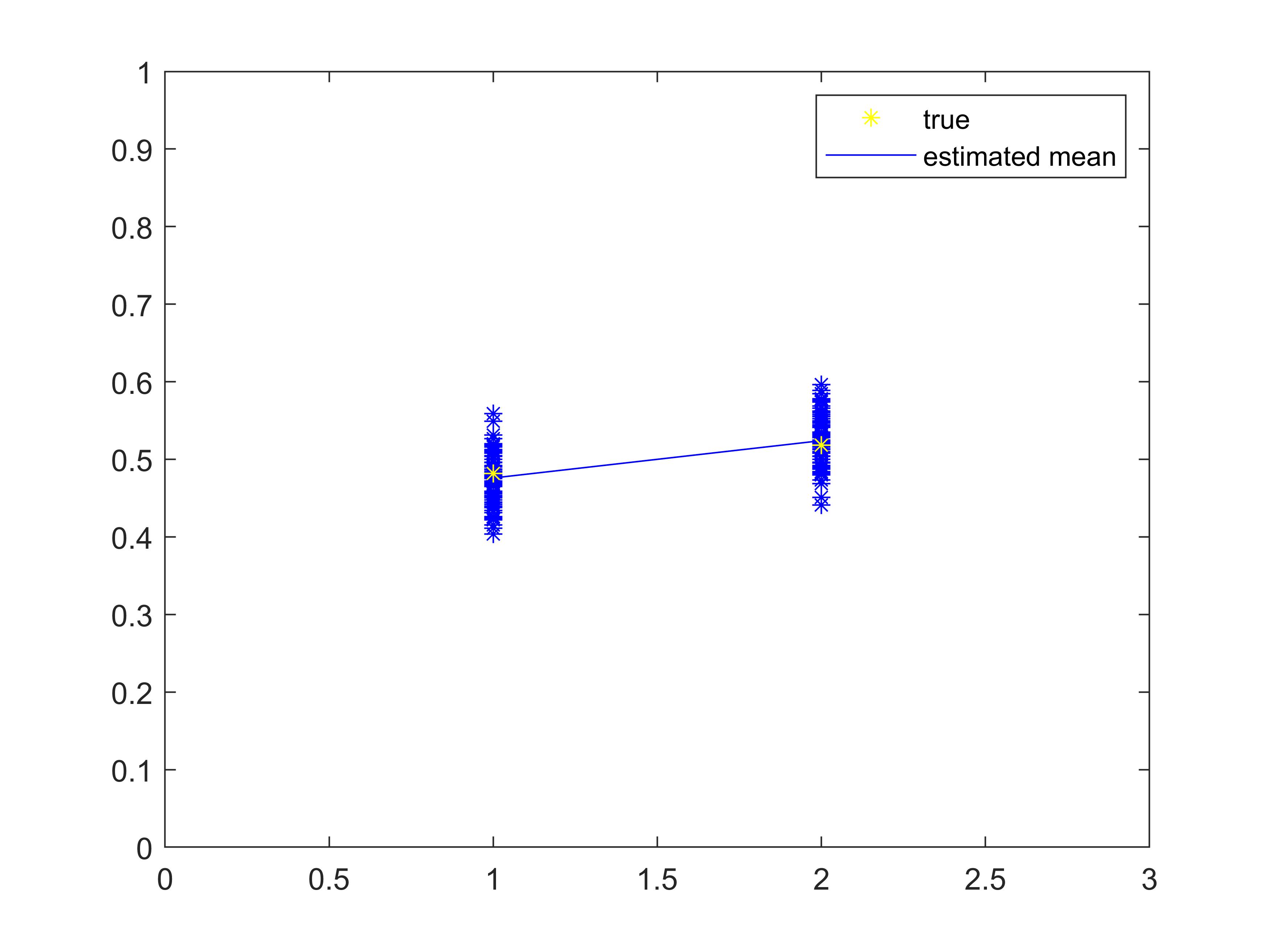}}
\subfigure[]{
\includegraphics[width=0.40\linewidth]{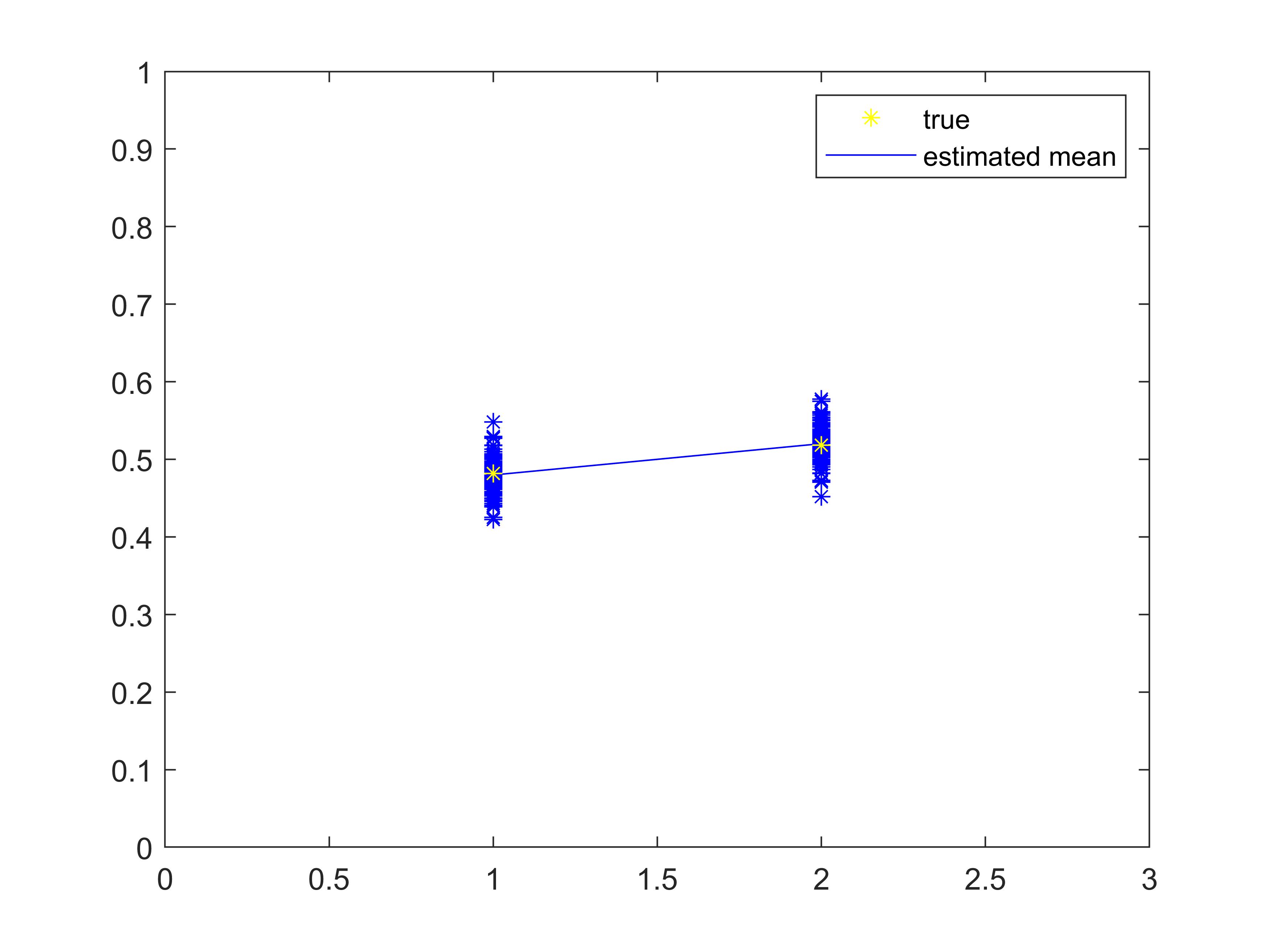}}\\
\caption{NY dataset: estimates of the true probabilities generating the data. The $x$-axis encodes the $S=2$ possible categories (female or male), for each one the yellow point represents the true probability $p_k$, while the solid blue line connects the estimated probabilities averaged over $100$ iterations.}\label{fig:ny}
\end{figure}

\begin{figure}[h!]
\begin{center}
{\includegraphics[width=0.40\linewidth]{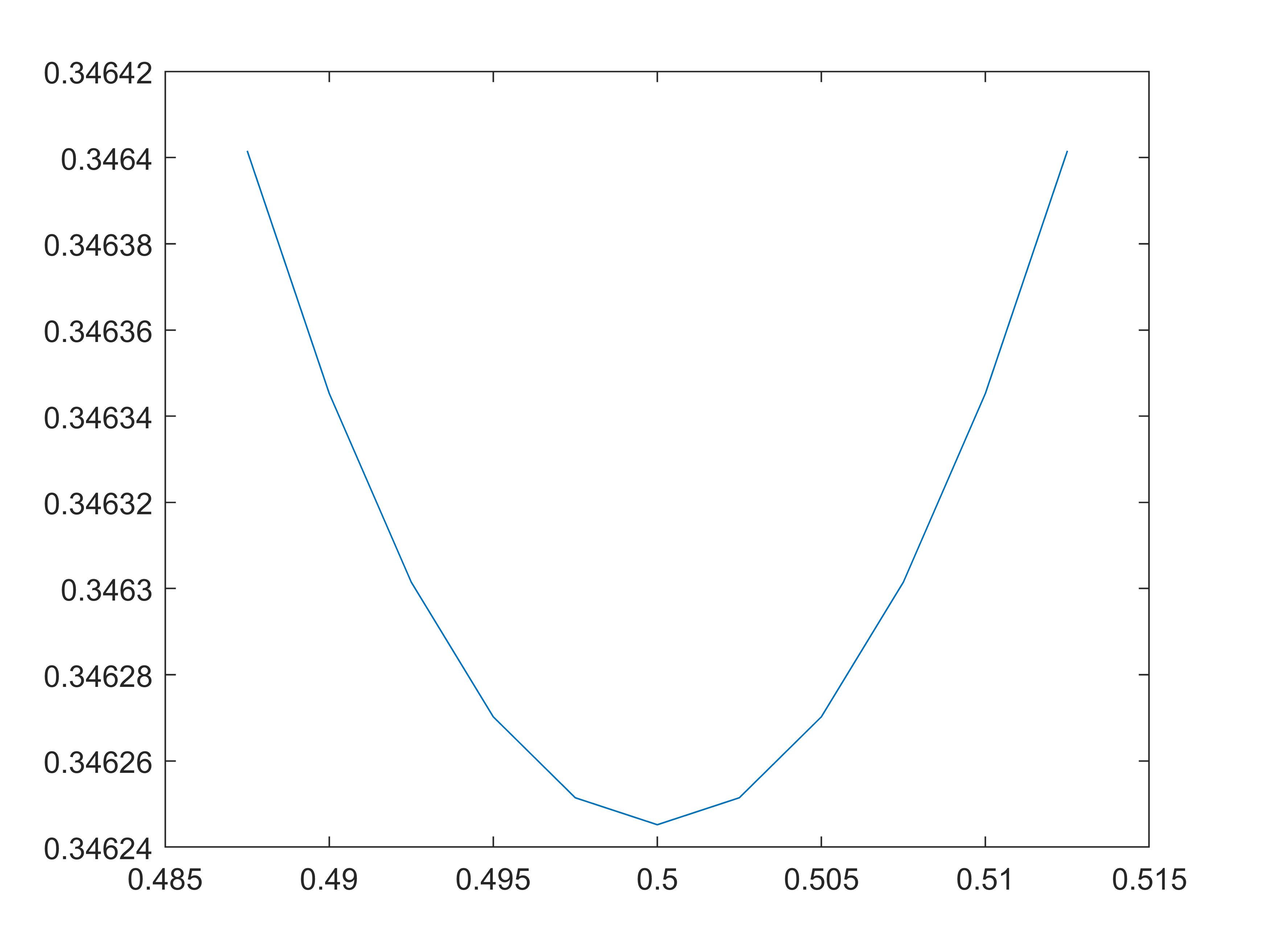}}   
\end{center}
\caption{NY dataset: mutual information as a function of $q$.}\label{fig:information}
\end{figure}

We finally test the performance of our strategy on some benchmark datasets from the public use microdata sample of the U.S. 2000 census for the state of New York, \cite{Rug(10)}.
The data contains the values of ten categorical variables of  $953076$ individuals: ownership of dwelling ($3$ levels),
mortgage status ($4$ levels), age ($9$ levels), sex ($2$ levels), marital status ($6$ levels),
single race identification ($5$ levels), educational attainment ($11$ levels), employment status ($4$ levels), work disability status ($3$ levels), and veteran status ($3$ levels).\\
For ease of illustration we consider the sex variable, which has two possible categories (female or male), therefore $S=2$ and $q=q_1=q_2$. We have already seen that the optimal $q$ lies on the boundaries of the interval 
$J_\alpha := [1/(e^\alpha +1), e^\alpha /(1+e^\alpha)]$.
We have estimated the probabilities of the two possible categories using the sample mean, thus obtaining $p_1=0.48$ and $p_2=0.52$.
In our numerical experiments we have considered $\alpha=0.05$, and for different values of $q \in J_\alpha$ we have generated the privatized dataset $Z_{1:n}$ estimating $p_1$ and $p_2$. More precisely, in Figure \ref{fig:ny} we have considered six values of $q \in [0.4875,  0.5125]$, and we reported the estimates  of  $p_1$ and $p_2$ averaged over $100$ iterations. Each panel corresponds to a different $q$, 
each blue star corresponds to the estimated value in one of the $100$ experiments based on the privatized sample $Z_{1:n}$. The solid blue line links the averaged estimates of the $p_k$'s over the $100$ runs, while the true values of the probabilities are represented in yellow. From the top left to the bottom right, we have chosen $q=0.4875  , 0.4925   ,    0.4975  ,  0.5025 ,  0.5075    ,  0.5125$: from the theory developed in the paper it is not surprising to realize that the values on the boundary lead to more reliable estimates, indeed they maximize the mutual information between $X$ and $Z$. 
In Figure \ref{fig:information} we reported the estimated mutual information between $X$ and $Z$  for different values of $q \in J_{0.05}$, in order to do that
we have estimated  $P_Z(k)$  and $P_X(k)$ using the corresponding sample means for each $k=1,2$.

\begin{center}
Figures \ref{fig:ny}--\ref{fig:information} about here.
\end{center}

\section{Conclusions and future work}\label{sec:5}
In this work, we have proposed a novel approach to choose the randomizing matrix $M$ in PRAM. This approach applies popular tools from computer science to derive $M$ as the solution of a constrained optimization problem, in which the Mutual Information between raw and transformed data is maximized, under the constraint that the transformation satisfies Differential Privacy. The proposed approach has the advantage to be quick and easy to implement. Also, the desired level of privacy can be tuned by a single parameter $\alpha$. \\
There are different ways in which the present work could be extended. A first possible direction of research is to understand how to tune the Differential Privacy parameter $\alpha$, which regulates the desired level of privacy, using the classical measures of risk \eqref{riskindeces1} and \eqref{riskindeces2}. Specifically, given the choice of some model, $\alpha$ can be chosen in such a way that the estimate of the disclosure risk index computed on the transformed dataset matches or falls below a particular threshold value. A second direction of research is to generalize the proposed procedure to the case in which a few categorical variables are jointly perturbed.  The proposed methodology can be extended to this case following similar lines. In particular, an individual will be randomly moved from one frequency cell to another using a $K\times K$ stochastic matrix, where $K$ is the number of cells after cross-classifying the variables we want to jointly perturb. In this setting, it will be important to study what further structure the $K\times K$ matrix should have in order to avoid structural zeros combinations.
Further, another direction of research is to examine the problem using other formulations of Differential Privacy. Specifically, the definition of Differential Privacy as in \eqref{diff priv} is known to provide a very strong privacy guarantee. Therefore, generalizing the proposed methodology to other formulations and relaxations of Differential Privacy, as those mentioned in Subsection \ref{Diff Priv}, might be an interesting topic. \\
Some other important research directions have also been suggested by the reviewers. Specifically, the proposed approach is focused on perturbing categorical variables, which are usually the most sensitive in terms of disclosure risk. However, a direction of research can be to study the problem for other datatypes, possibly including also some continuous data. 
Another line of research is to study theoretical guarantees in terms of preservation of utility for different classes of queries. In computer science, with a query it is usually meant a statistics of the observations, or function of some sufficient statistics. A crucial problem consists to quantify and analyse the expected distance (risk) of some classes of queries computed on raw and realised dataset. Similar contributions in this direction are \cite{Smi11} and \cite{Duc18}.
A useful extension of the proposed methodology would focus on different structures for the matrix $M$, rather than with uniform off-diagonal rows as in \eqref{matrix_form}. An interesting example of application
suggested by one reviewer, in which imposing non-uniform off-diagonal rows would be important,  is
in spatial modelling, when perturbing a geographical variable.
In this context, a more suitable structure for $M$ would allow for the geographical category to have a higher probability to be swapped with a spatially neighboring category rather than to one very far from the true observed value.
Within this context, the optimal choice of $M$ will have to balance between the higher randomization to achieve the same level of Differential Privacy and the benefit in statistical utility that follows from geographically localised perturbation for any later spatial analysis. 
Finally, another extension could be to include all variables in the mutual information in the maximization \eqref{Maximization}, applying privacy perturbation and the Differential Privacy constraint only to a subset of them. If the included and excluded variables are modelled as independent, the solution of the maximization problem $M$ should be unaltered. Instead, in the dependent case, the optimal solution $M$ might depend also on the non-perturbed variables and the maximization problem could become analytically much more challenging.

\section*{Aknowlegdment}

The authors thank the Associate Editor and two anonymous referees, whose constructive comments and suggestions have been appreciated and helped to improve the paper. Federico Camerlenghi received funding from the European Research Council (ERC) under the European Union’s Horizon 2020 research
and innovation programme under grant agreement No 817257. Federico Camerlenghi gratefully acknowledge the financial support from the Italian Ministry of Education, University and Research (MIUR), ``Dipartimenti di Eccellenza'' grant 2018-2022.

\section*{Appendix} \label{Appendix}

\subsection{Proof of Proposition \ref{prop_optimal_max}}
To underline the dependency on $q$, we will sometimes use the notation $Q_q$. We need to show that $f$ is convex. Let $q' = (q'_1,..,q'_S)^T$ and $\theta \in [0,1]$. Let $k,l \in \{1,..,S\}$ such that $k\not = l$.
\begin{align*}
&Q_{\theta q + (1-\theta)q'}(Z=k|X=k) = \theta q_k + (1-\theta)q_k' \\
&\qquad =\theta Q_q(Z=k|X=k ) + (1-\theta)Q_{q'}(Z=k|X=k).
\end{align*}
Besides,
\begin{align*}
&Q_{\theta q + (1-\theta)q'}(Z=l|X=k) \\
& \qquad = \frac{1- (\theta q_k + (1-\theta)q_k')}{S-1} \\
&\qquad=\frac{\theta(1- q_k) + (1-\theta)(1-q_k')}{S-1} \\
&\qquad= \theta Q_q(Z=l|X=k) + (1-\theta)Q_{q'}(Z=l|X=k).
\end{align*}
Therefore, $Q_{\theta q + (1-\theta)q'} = \theta Q_q + (1-\theta)Q_{q'}$. It is known that for a fixed marginal distribution of one of the variables, the mutual information is convex in the conditional distribution of the second, see for example Theorem 2.7.4 of \citet{Cov12}. Therefore, 
$f(\theta q + (1-\theta)q') \leq \theta f(q) + (1-\theta)f(q'),$
and hence $f$ is convex.

\subsection{Fact 1: Set of feasible parameters $q$}


We start by writing explicitly the linear constraints \eqref{linear_constraints} on $q$. Let $\mathcal{T}_\alpha$ be the convex polytope of all $q$ satisfying $\alpha$-differential privacy. Let $\mathcal{S}_\alpha$ be the planar polygon defined by the set of equations
\begin{eqnarray}
(S-1)x+e^\alpha y &\leq & e^\alpha \label{constraint1}\\
(S-1)y+e^\alpha x &\leq & e^\alpha \label{constraint2}\\
-(S-1)e^\alpha y - x &\leq & -1 \label{constraint3}\\
-(S-1)e^\alpha x + y &\leq & -1 \label{constraint4}\\
e^\alpha y - x &\leq & e^\alpha - 1 \label{constraint5}\\ 
e^\alpha x - y &\leq & e^\alpha - 1 \label{constraint6}
\end{eqnarray}
The set of feasible points is then characterized by 
\begin{equation*}
    (q_1,\dots,q_S) \in \mathcal{T}_\alpha \iff \forall (k,l),\ (q_k,q_l) \in \mathcal{S}_\alpha. \label{characterisation_T}
\end{equation*}
This set  characterized by the $3S(S-1)$ linear constraints given by equations \eqref{constraint1} to \eqref{constraint6} can thus be defined as the set of solutions of the equation $C q^T \leq b_\alpha$, where $C$ has dimension $3S(S-1) \times S$ and $b_\alpha$ is a $3S(S-1)$-dimensional vector.  

\subsection{Proof of Proposition \ref{prop_feasible_set}}

Equations \eqref{constraint1} to \eqref{constraint4} define a quadrilateral whose vertices are 
\begin{eqnarray*}
u_{\alpha} &=& \left( \frac{(S-1)e^\alpha-1}{S(S-2)}, \frac{(S-1)e^{-\alpha}-1}{S(S-2)} \right), \\
u_{-\alpha} &=& \left( \frac{(S-1)e^{-\alpha}-1}{S(S-2)} , \frac{(S-1)e^{\alpha}-1}{S(S-2)} \right), \\
v_{\alpha} &=& \left( \frac{e^{\alpha}}{S-1+e^{\alpha}}, \frac{e^{\alpha}}{S-1+e^{\alpha}} \right), \\
v_{-\alpha} &=& \left( \frac{e^{-\alpha}}{S-1+e^{-\alpha}}, \frac{e^{-\alpha}}{S-1+e^{-\alpha}} \right).
\end{eqnarray*}
The points $v_{\alpha}$ and $v_{-\alpha}$ always satisfy \eqref{constraint5} and \eqref{constraint6}. Besides, for $S \geq 4$, if $\alpha \leq \log (S+\sqrt{S(S-4)}-2)-\log 2$, then $u_\alpha$ and $u_{-\alpha}$ also satisfy \eqref{constraint5} and \eqref{constraint6}. In such a setting, equations \eqref{constraint5} and \eqref{constraint6} are redundant and hence can be omitted when defining $\mathcal{T_{\alpha}}$. Common values of $\alpha$ are generally within the range $[0,2]$, therefore equations \eqref{constraint5} and \eqref{constraint6} are omitted when $S \geq 10$. In the following, we will suppose that $S\geq 4$ and $\alpha \leq \log (S+\sqrt{S(S-4)}-2)-\log 2$.
Let $(q_1,\dots, q_S) \in \mathcal{T}_\alpha$, since $(q_2, q_3)$ satisfy \eqref{constraint4}, we have that $q_3 \leq 1-(S-1)e^\alpha q_2$. Therefore, using the fact that $(q_1, q_3)$ satisfy \eqref{constraint1}, and the symmetry of the constraints, we can deduce that any $(q_k, q_l)$ satisfy
\begin{eqnarray}
y-e^{2\alpha}x \geq 1 \label{constraint7} \\
x-e^{2\alpha}y \geq 1 \label{constraint8}
\end{eqnarray}
Equations \eqref{constraint2} and \eqref{constraint7} give that $q_k \leq \frac{e^{\alpha}}{e^{-\alpha}+S-1} = v_{\max}$ and equations \eqref{constraint4} and \eqref{constraint7} give $q_k \geq \frac{e^{-\alpha}}{e^{\alpha}+S-1} = v_{\min}$. Let $\mathcal{V}$ be the set of points defined up to permutations by  
\begin{enumerate}
    \item $\forall k,\, q_k = \frac{e^{\alpha}}{e^{\alpha}+S-1} = v_\alpha$
    \item $\forall k,\, q_k  = \frac{e^{-\alpha}}{e^{-\alpha}+S-1} = v_{-\alpha}$
    \item $ \forall k,\, q_{k} = v_{i_k \alpha}$ with $i_k = \pm 1$ and $2 \leq  \#  \{k\ \text{s.t } i_k = 1\} \leq S-2$
    \item $q_1 = v_{\min}$, $q_{k \geq 2} = v_\alpha$
   \item $q_1 = v_{\max}$, $q_{k\geq 2} = v_{-\alpha}$. 
\end{enumerate}
Under the assumption that $S \geq 4$ and $\alpha \leq \log (S+\sqrt{S(S-4)}-2)-\log 2$, it is straightforward to verify that $\mathcal{V} \subset \mathcal{T}_\alpha$. In the following, we show that any element of $\mathcal{T}_\alpha$ is a convex combination of points of $\mathcal{V}$. In order to do so, we will use the following Lemma.

\begin{lemma}
For $S \geq 2$, if $q$ satisfies differential privacy, then at most one of its coordinates is larger than $v_\alpha$ and at most one is smaller than $v_{-\alpha}$
\end{lemma}

\textbf{Proof}
This trivially follow from the constraint \eqref{max_constrained}. Indeed, suppose that $q_{i} > \frac{e^{\alpha}}{S-1+e^\alpha}$. Then for any other $q_{j}$, using formula (19),

\[
 (S-1)\frac{e^{\alpha}}{S-1+e^\alpha}+e^{\alpha}q_{j} < (S-1)q_{i}+e^{\alpha}q_{j}\leq e^\alpha
 \]
Therefore,
\[
 (S-1)\frac{1}{S-1+e^\alpha}+q_{j}  < 1
\]
\[ q_{j} < 1- (S-1)\frac{1}{S-1+e^\alpha}=\frac{e^{\alpha}}{S-1+e^\alpha}
\]
Similarly suppose both $q_{i} < \frac{e^{-\alpha}}{S-1+e^{-\alpha}}$. For any other $q_{j}$, from formula (21), 
\[
(S-1)e^{\alpha}\frac{e^{-\alpha}}{S-1+e^{-\alpha}}+q_{j} > (S-1)e^{\alpha}q_{i}+q_{j}\geq 1
\]
Therefore
\[
\frac{(S-1)}{S-1+e^{-\alpha}}+q_{j} >  1
\]
\[q_{j} >  \frac{e^{-\alpha}}{S-1+e^{-\alpha}}\]
\qed

Let $(q_1,\dots, q_S) \in \mathcal{T}_\alpha$, using previous Lemma, we know that up to permutations, one of 4 settings is possible:
\begin{enumerate}
    \item For all $k$, $\ v_{-\alpha} \leq q_k \leq v_\alpha$. 
   \item $v_{\alpha} < q_1 \leq  \frac{e^{\alpha}}{e^{-\alpha}+S-1} = v_{\max}$, and for 
   $ k \geq 2$, $v_{-\alpha} \leq q_k \leq v_{\alpha}$
    \item $ v_{\min} = \frac{e^{-\alpha}}{e^{\alpha}+S-1} \leq q_1 < v_{-\alpha}$, and for $ k \geq 2$, $v_{-\alpha} \leq q_k \leq v_\alpha$
    \item $v_\alpha < q_1 \leq  v_{\max}$, $ v_{\min} \leq q_2 < v_{-\alpha}$, and for $ k \geq 3$, $v_{-\alpha} \leq q_k \leq v_\alpha$
\end{enumerate}

\vspace{1em}
The first setting is the most straightforward, indeed since $v_{\min} < v_{-\alpha}$ and  $v_{\max} > v_{\alpha}$, we find that all the points $(v_{i_k \alpha})_{1\leq k \leq S}$ for any sequence $(i_k)_{1\leq k \leq S} \in \{ -1,1\}^S$, are within the convex hull of $\mathcal{V}$ and so does the whole hypercube generated by those $2^S$ points.

The second and third settings have similar proof, that we will explicit for the second setting. As said in the previous remark, the point $(v_\alpha, v_{-\alpha}, \dots, v_{-\alpha})$ belongs to the convex hull of $\mathcal{V}$. Let $k \geq 2$, we know that $q_k \geq v_{-\alpha}$. Besides, since $(q_k, q_1) \in \mathcal{S}_\alpha$, \eqref{constraint1} gives that $(q_k, q_1)$ is below the line passing through $(v_{-\alpha}, v_{\max})$ and $(v_\alpha, v_\alpha)$. Hence, denoting $\theta = \frac{q_1-v_\alpha}{v_{\max}-v_\alpha}$, we find that
\[
v_{-\alpha} \leq q_k \leq \theta v_{-\alpha} + (1-\theta) v_{\alpha}.
\]
Therefore, we only need to show that any point\\
$(q_1, x_2, \dots, x_{S})$ is in the convex hull of $\mathcal{V}$ for any sequence $(x_k)_{k\geq 2} \in \{v_{-\alpha},\   \theta v_{-\alpha} + (1-\theta) v_{\alpha}\}^{S-1}.$ Let $(q_1, x_1, \dots, x_{S-1})$ be such a point. Let $(i_k)_{2\leq k \leq S}$ such that $i_k = -1$ if $x_k = v_{-\alpha}$, and $i_k = 1$ otherwise. Now, from previous setting we know that $(v_\alpha, v_{i_2 \alpha},\dots, v_{i_S \alpha})$ is in the convex hull of $\mathcal{V}$, and so does $(v_{\max},v_{-\alpha},\cdots, v_{-\alpha}).$ We conclude as we notice that 
\begin{align*}
 &\theta (v_{\max},v_{-\alpha},\cdots, v_{-\alpha}) + (1-\theta) (v_\alpha, v_{i_2 \alpha},\dots, v_{i_S \alpha}) \\
 & \qquad= (q_1, x_2, \dots, x_{S})
\end{align*}

The proof of the last setting is similar to the previous one. Equation \eqref{constraint8} together with $x \geq v_\alpha$ and $y \leq v_{-\alpha}$ define a triangle within which $(q_1, q_2)$ lays. The points $(v_{\max}, v_{-\alpha}), (v_{\alpha}, v_{\min})$ and $(v_\alpha, v_{-\alpha})$ are the three vertices of the triangle. Therefore, denoting $\theta_1 = \frac{q_1-v_\alpha}{v_{\max}-v_\alpha}$ and $\theta_2 = \frac{v_{-\alpha}-q_2}{v_{-\alpha}-v_{\min}}$, we have that $0 \leq \theta_1, \theta_2 \leq 1$, $\theta_1+\theta_2 \leq 1$ and $(q_1, q_2)$ equals
\[ \theta_1 (v_{\max}, v_{-\alpha}) + \theta_2 (v_{\alpha}, v_{\min}) + (1-\theta_1 - \theta_2) (v_\alpha, v_{-\alpha}).\]
Let $k \geq 3$, since $(q_k, q_1) \in \mathcal{S}_\alpha$, \eqref{constraint1} implies that $(q_k, q_1)$ is below the line passing through $(v_{-\alpha}, v_{\max})$ and $(v_\alpha, v_\alpha)$. Similarly, since $(q_2, q_k) \in \mathcal{S_\alpha}$, \eqref{constraint4} implies that $(q_2, q_k)$ is above the line passing through $(v_{\min}, v_\alpha)$ and $(v_{-\alpha}, v_{-\alpha})$. Therefore, $q_k$ satisfies
\[ \theta_2 v_{\alpha} + (1-\theta_2) v_{-\alpha} \leq q_k \leq \theta_1 v_{-\alpha} + (1-\theta_1)v_\alpha.\]
As previously, we only need to show that any point $(q_1, q_2, x_3, \dots, x_S)$ is in the convex hull of $\mathcal{V}$ for any $(x_k)_{k\geq 3} \in \{\theta_2 v_{\alpha} + (1-\theta_2) v_{-\alpha},\  \theta_1 v_{-\alpha} + (1-\theta_1) v_{\alpha}\}^{S-2}.$ Let $(x_k)_{k \geq 3}$ be such a sequence. Let $(i_k)_{k \geq 3}$ defined by $i_k = -1$ if $x_k = \theta_2 v_{\alpha} + (1-\theta_2) v_{-\alpha}$ and $i_k = 1$ otherwise. We conclude by noticing that 
\begin{align*}
& (q_1, x_2, \dots, x_{S}) \\
& \qquad=\theta_1 (v_{\max},v_{-\alpha},\cdots, v_{-\alpha}) +\theta_2 (v_{\alpha},v_{\min},v_{\alpha},\cdots, v_{\alpha})\\
& \qquad \qquad + (1-\theta_1 - \theta_2) (v_\alpha, v_{-\alpha}, v_{i_3 \alpha},\dots, v_{i_s \alpha}).
\end{align*}

\clearpage

\clearpage

\clearpage
\clearpage

\end{document}